\documentclass[journal, twoside]{IEEEtran}
\usepackage{subcaption}
\usepackage{indentfirst}
\usepackage{graphicx}
\usepackage{amssymb}
\usepackage{amsfonts}
\usepackage{mathrsfs}
\usepackage{leftidx}
\usepackage{color}
\usepackage{amsmath}
\usepackage{arydshln}
\usepackage{amsthm}
\usepackage{ragged2e}
\usepackage{cite}
\usepackage{enumerate}
\usepackage{longtable}
\usepackage{float}
\usepackage{stfloats}
\usepackage{hyperref}
\usepackage{algpseudocode}
\usepackage{algorithm}
\usepackage{makecell}
\usepackage{url}
\usepackage{enumitem}
\usepackage{multirow} 
\usepackage{booktabs}
\theoremstyle{plain}

\theoremstyle{remark}
\newtheorem{rem}{Remark}

\usepackage{caption}
\usepackage{makecell}

\newcolumntype{P}[1]{>{\raggedright\arraybackslash\footnotesize}m{#1}}
\newcolumntype{A}[1]{>{\centering\arraybackslash\footnotesize}m{#1}}

\usepackage[table,usenames,dvipsnames]{xcolor}
\definecolor{aa}{RGB}{175,238,238}
\definecolor{bb}{RGB}{255,255,255}

\usepackage{bm}
\usepackage{makecell}

\begin{document}

\title{SemSteDiff: Generative Diffusion Model-based Coverless Semantic Steganography Communication}

\author{Song Gao, Rui Meng,~\IEEEmembership{Member,~IEEE,} Xiaodong Xu,~\IEEEmembership{Senior Member,~IEEE,} 
Haixiao Gao, 

Yiming Liu,~\IEEEmembership{Member,~IEEE,} Chenyuan Feng,~\IEEEmembership{Member,~IEEE,} 
Ping Zhang,~\IEEEmembership{Fellow,~IEEE,} 

Tony Q. S. Quek,~\IEEEmembership{Fellow,~IEEE,} and Dusit Niyato,~\IEEEmembership{Fellow,~IEEE}

\thanks{
This work was supported in part by the National Key Research and Development Program of China under Grant 2020YFB1806905; in part by the National Natural Science Foundation of China under Grant 62501066 and under Grant U24B20131; in part by the Beijing Municipal Natural Science Foundation under Grant L242012; in part by the Long Term Science and Technology Plan for Broadcasting, Television, and Online Audiovisual Programs under Grant 2025AD0300; and in part by the National Research Foundation, Singapore and Infocomm Media Development Authority under its Communications and Connectivity Bridging Funding Initiative. Any opinions, findings and conclusions or recommendations expressed in this material are those of the author(s) and do not reflect the views of National Research Foundation, Singapore.
\textit{(Corresponding authors: Rui Meng and Tony Q. S. Quek.)}

Song Gao, Rui Meng, Xiaodong Xu, Haixiao Gao, Yiming Liu, and Ping Zhang are with State Key Laboratory of Networking and Switching Technology, Beijing University of Posts and Telecommunications, Beijing, China (e-mail: wkd251292@bupt.edu.cn; buptmengrui@bupt.edu.cn; xuxiaodong@bupt.edu.cn; haixiao@bupt.edu.cn; liuyiming@bupt.edu.cn; pzhang@bupt.edu.cn).

Chenyuan Feng is with Department of Computer Science, University of Exeter, EX4 4QF Exeter, U.K. (e-mail: c.feng@exeter.ac.uk).

Tony. Q. S. Quek is with the Singapore University of Technology and Design, Singapore 487372, and also with the Department of Electronic Engineering, Kyung Hee University, Yongin 17104, South Korea (e-mail: tonyquek@sutd.edu.sg).

Dusit Niyato is with College of Computing and Data Science, Nanyang Technological University, Singapore (e-mail: dniyato@ntu.edu.sg).

}}

\maketitle

\begin{abstract}
Semantic communication (SemCom), as a novel paradigm for future communication systems, has recently attracted much attention due to its superiority in communication efficiency. However, similar to traditional communication, it also suffers from eavesdropping threats. Intelligent eavesdroppers could launch advanced semantic analysis techniques to infer secret semantic information. 
Therefore, some researchers have designed Semantic Steganography Communication (SemSteCom) schemes to confuse semantic eavesdroppers. However, the state-of-the-art SemSteCom schemes for image transmission rely on the pre-selected cover image, which limits the generalization.
To address this issue, we propose a Generative Diffusion Model-based Coverless Semantic Steganography Communication (SemSteDiff) scheme to hide secret images into generated stego images. The semantic related private and public keys enable legitimate receiver to decode secret images correctly while the eavesdropper without the completely correct key-pairs fail to obtain them. Simulation results demonstrate the effectiveness of the plug-and-play design in different Joint Source-Channel Coding (JSCC) frameworks. 
Results under different eavesdropping settings show that, when Signal-to-Noise Ratio (SNR) = 0 dB, the peak signal-to-noise ratio (PSNR) of the legitimate receiver is 4.14 dB higher than that of the eavesdropper.


\end{abstract}


\section{Introduction}


\subsection{Background}
With the deep convergence of Artificial intelligence (AI) and communication technologies, the sixth-generation mobile communication system (6G) is expected to connect people, machines, things, and intelligence together to achieve “Internet of Intelligence”. As a pivotal direction for 6G, \textit{intellicise (intelligent and concise)} wireless network, supported by semantic communication (SemCom), can enable a more intelligent, efficient, and low-complexity system \cite{meng2026intellicise}. In contrast to traditional communication, SemCom aims to transmit the meaning of messages, called semantic features, rather than transmitting specific bits, thus significantly improving communication efficiency and reducing bandwidth requirements \cite{10854543,cheng2026apeg}. 

Numerous researchers have devoted their efforts to SemCom and proposed various key technologies to facilitate its application in 6G, such as generative AI-aided semantic successive refinement \cite{zhang2025semantic}, semantic-forward relaying \cite{lin2024semantic}, and adaptive NOMA-based interference-aware transmission \cite{yan2024adaptive}. For instance, Lin \textit{et al.} \cite{lin2024semanticIC} investigated semantic interference cancellation for wireless networks, demonstrating that semantic-level signal processing can effectively mitigate co-channel interference and enhance transmission reliability in multi-user 6G scenarios. While SemCom is vulnerable to eavesdropping threats due to the openness of wireless channels \cite{shen2023secure}, making security becomes a significant topic. Signals transmitted over channel can be intercepted by unauthorized receivers operating within the same frequency band. Once eavesdroppers gain access to semantic decoders, they could further obtain the private information even under poor channel conditions \cite{meng2025survey}. Therefore, how to defend against semantic eavesdroppers has become a significant issue.

\subsection{Secure SemCom}

Encryption techniques are important for privacy protection in traditional communication systems. Motivated by this,
Tung et al.\cite{tung2023deep} proposed the first secure SemCom scheme to resist eavesdropping attacks. It leverages the affine property of a public-key-encryption scheme based on learning with errors. Some researchers further propose secure SemCom schemes based on physical-layer Advanced Encryption Standard (AES) encryption \cite{chen2024lightweight}, homomorphic encryption \cite{meng2025secure}, and quantum encryption \cite{khalid2023quantum}.
Additionally, beamforming \cite{dai2024secure}, reconfigurable intelligent surface \cite{zhao2022semkey}, and adversarial training \cite{luo2023encrypted} are also introduced to enhance the security of SemCom.

Besides cryptography-based encryption techniques, researchers also explore covert communications for SemCom, which aim to defend against eavesdroppers by concealing the communication behavior between legitimate transmitters and receivers. Wang et al.\cite{wang2023multi} combined multi-agent reinforcement learning to enable each device and jammer to collaborate in identifying vulnerable eavesdroppers, thereby deriving strategies that jointly maximize semantic information transmission and power control. Also, Xu et al.\cite{xu2024covert} designed a covert SemCom framework for Unmanned Aerial vehicle (UAV) scenarios by jointly optimizing flight trajectory and transmission power. Furthermore, Liu et al.\cite{liu2025learning} proposed a covert SemCom system that supports multiple modalities, including text, images, and audio. It employs a power control method, which not only ensures the efficiency of covert communication but also achieves high-quality semantic decoding.

Inspired by covert communication, researchers have recently studied Semantic Steganography Communication (SemSteCom) by introducing steganography technology into SemCom
\cite{long2025scf,li2024multi,tang2025towards,tang2024secure,huo2025image,chencoding,meng2025image}. By embedding the secret information into cover information, steganography provides a new defense method for SemCom to achieve the target of “invisible encryption”. Long et al.\cite{long2025scf} introduced a Linguistic steganography scheme, which enhances text imperceptibility and semantic coherence using a knowledge graph to guide secret encoding. Li et al.\cite{li2024multi} proposed a multi-modal steganography scheme that hides texts into images to achieve secure SemCom. Towards image transmission, Tang et al.\cite{tang2025towards} proposed an Invertible Neural Network (INN)-based SemSteCom scheme, which covertly embeds the semantic representation of a private image into the channel input of a host image, yet the eavesdropper can only detect the host image. Besides, Huo et al.\cite{huo2025image} constructed a Generative Adversarial Networks (GAN)-based SemSteCom scheme, considering both pixel-level and semantic-level distortions to enhance SemCom security. Also, Chen et al.\cite{chencoding} embedded protected information into the transmission process through a coding-guided jamming strategy. It adapts a two-layer superposition coding structure to prevent eavesdroppers from extracting the privacy source contents.

\begin{table*}[htpb]
\centering
\small
\caption{Comparison between existing SemSteCom schemes and the proposed scheme, where modality refers to the source of the transmitted information, framework refers to the core network structure that semantic steganography relies on, classification refers to whether the steganographic scheme uses cover images, eavesdropper refers to whether the proposed scheme considers the eavesdropping situation, and the evaluation refers to the performance metrics.
}
\begin{tabular}{@{}l c l  c c l @{}}
\toprule
\textbf{Reference} & \textbf{Modality} & \textbf{Framework} &  \textbf{Classification}  & \textbf{Eavesdropper} & \textbf{Evaluation Metrics} \\
\midrule

\cite{long2025scf} & Text & Knowledge graph & Coverless & No & BLEU, METEOR, ROUGE\\

\cite{li2024multi} & Text & Invertible Neural Network &  Cover-edited & Yes & Overall Accuracy \\

\cite{tang2025towards,tang2024secure} & Image & Invertible Neural Network & Cover-edited & Yes & PSNR, SSIM, LPIPS \\

\cite{huo2025image} & Image & Generating Adversarial Network & Cover-edited & No & PSNR, SSIM, Accuracy \\

\cite{chencoding} & Image & Coding-Enhanced jamming & Cover-edited & Yes & PSNR \\

Ours & Image & Generative Diffusion Model & Coverless & Yes & PSNR, SSIM, LPIPS, MSE \\
\bottomrule
\end{tabular}
\label{tab:stegano_comparison}
\end{table*}

\subsection{Motivations and Contributions}
Although state-of-the-art SemSteCom schemes embed secret images into cover images to deceive eavesdroppers, they still suffer from the following limitations. First, the hiding of secret images is constrained by the size and quality of the cover images. This implies that transmitting a high-dimensional secret image requires a cover image with sufficiently high resolution and semantic characteristics\cite{zhou2022secret}. Second, the cover-edited method faces the robustness risk. If the cover image suffers from channel noise, compression, and non-linear transformations, it could damage the information of restored secret image\cite{yu2023cross}. Moreover, the embedding of hidden semantic features could cause statistical inconsistency. Once eavesdroppers employ advanced steganalysis techniques, such potential anomalies may serve as critical cues for detecting semantic steganography \cite{song2024survey}.

\begin{figure}[]
    \centering
    \begin{subfigure}{0.5\textwidth}
        \includegraphics[width=\textwidth]{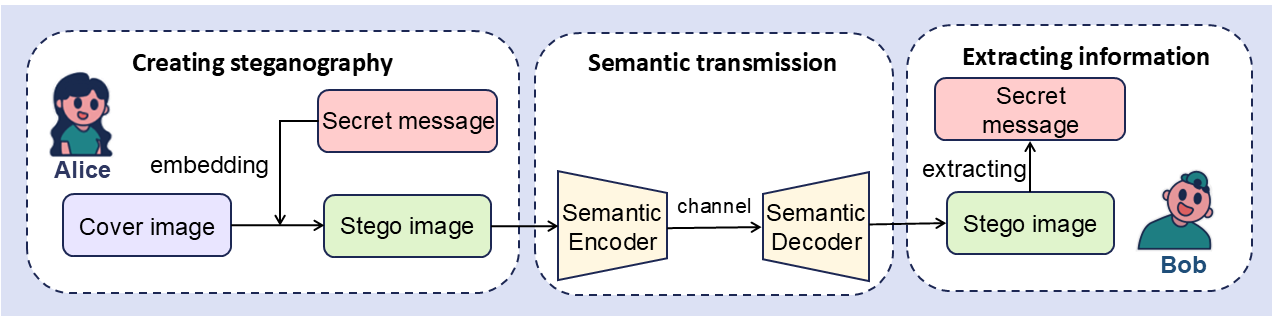}
        \caption{Existing cover-edited SemSteCom schemes for image transmission\cite{tang2025towards,tang2024secure,huo2025image,chencoding}.}
        \label{fig:A}
    \end{subfigure}
    \hfill
    \begin{subfigure}[b]{0.5\textwidth}
        \includegraphics[width=\textwidth]{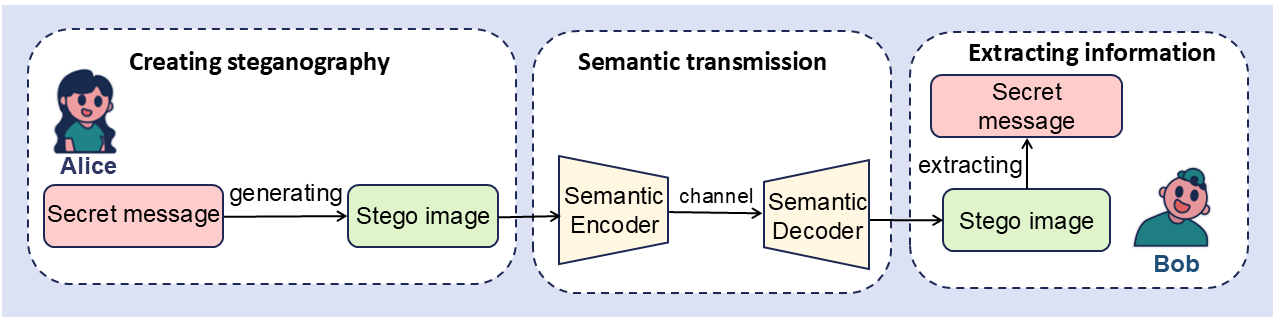}
        \caption{The proposed coverless SemSteDiff scheme for image.}
        \label{fig:B}
    \end{subfigure}
    \caption{Comparison between existing SemSteCom schemes and the proposed SemSteDiff scheme. (a) introduces the existing cover-edited SemSteCom schemes, which depend on embedding secret messages into cover images to obtain stego images. (b) introduces proposed coverless SemSteDiff scheme, which does not rely on cover images and generates stego images directly.}
    \label{fig:AB}
\end{figure}

To overcome the above challenges, we introduce generative diffusion models to realize coverless image SemSteCom. 
Generative diffusion models realize image generation through forward diffusion and backward denoising processes. This reversibility enables the secret image to be directly embedded in the generation process of stego image without relying on the pre-selected cover image. Additionally, the generation process can be controlled by conditional input, thus guaranteeing the controllability of stego images \cite{fan2025generative}. The comparison between existing SemSteCom schemes and the proposed Generative Diffusion Model-based Coverless Semantic Steganography Communication (SemSteDiff) scheme is illustrated in Table \ref{tab:stegano_comparison}.
The main contributions are described as follows:
\begin{itemize}
\item We propose the SemSteDiff scheme to defend against semantic eavesdroppers. It mainly consists of three plug-and-play modules, including the private and public key generation, conditional diffusion model-based coverless steganography, and Joint Source-Channel Coding (JSCC)-based semantic codec modules. As illustrated in Figure \ref{fig:AB}, compared with existing cover-edited image SemSteCom schemes, SemSteDiff overcomes the limitation of cover images, further enhancing the concealment of SemCom.
\item Under the idea of semantics-as-key, we design a Bootstrapping Language-Image Pretraining (BLIP)-based private key extractor to obtain semantic descriptions of secret images as private keys. We further propose a Large Language Model (LLM)-based public key generator to produce public keys in-pairs, thus achieving cross-modal protection. Our method dynamically generates semantic keys specific to actual content descriptions, ensuring precise semantic alignment. Moreover, the potential eavesdroppers can only obtain the public keys that reflect the semantics of stego images. Such design can mislead potential eavesdroppers and significantly reduce the risk of suspicion.

\item We explore the conditional diffusion model for SemSteCom. By controlling both the forward diffusion and reverse denoising processes with semantic keys, we achieve reliable coverless stego image generation. In our framework, we introduced the Variational Autoencoder (VAE) to map secret image into latent space. It offers a compressed representation and effective computational way to apply the diffusion process. To achieve the conditional control, the attention mechanism embeds public and private keys into diffusion, ensuring that only legitimate receivers can decode the secret image.

\item Simulations results on open-source Stego260 datasets \cite{yu2023cross} demonstrate that SemSteDiff enables the legitimate receiver to accurately reconstruct private images, while semantic eavesdroppers without correct private keys lead to wrong results. This confirms the effectiveness of SemSteDiff in ensuring both semantic accuracy and communication security.
\end{itemize}

\section{System Model}

\subsection{Network Model}
\begin{figure*}[htbp]
\centering
\vspace{-10mm}
\includegraphics[width=\linewidth]{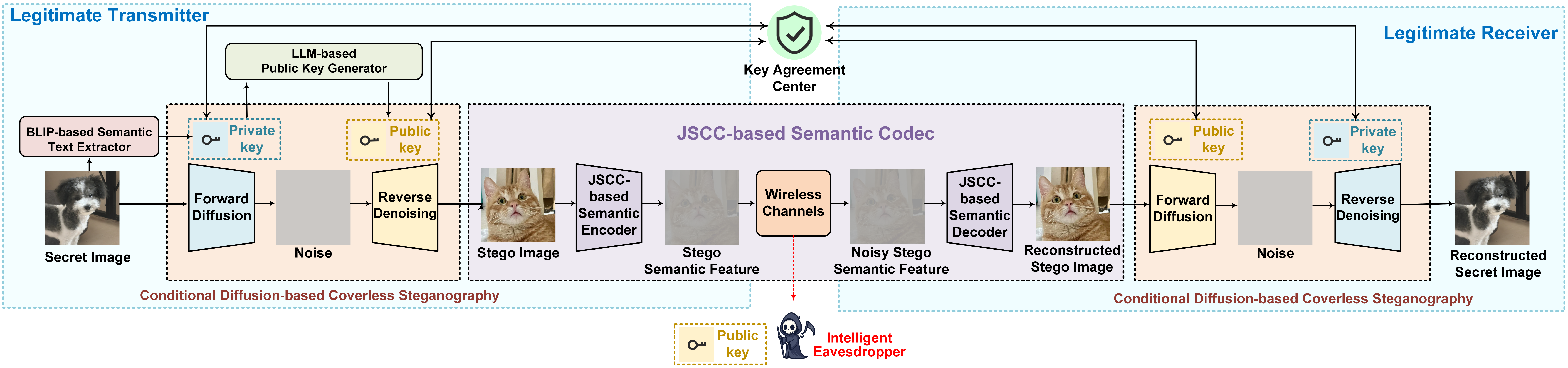}
\caption{The overview framework of SemSteDiff, where BLIP-based private key extractor obtains textual description of secret images as private keys, LLM-based public key generator produces public keys in pairs with private keys, conditional diffusion model-based coverless steganography module embeds keys' characteristic into attention mechanism to generate relative stego images, and JSCC-based semantic codec module achieves SemCom.}
\label{fig211}
\end{figure*}

As illustrated in Figure \ref{fig211}, we introduce SemSteDiff to achieve coverless and controllable SemSteCom. The involved nodes are described as follows.
\begin{itemize}
\item \textbf{Legitimate Transmitter and Receiver:} The legitimate transmitter uses private keys $K_{\text{priv}}$ and public keys $K_{\text{pub}}$ to hide secret images $\mathbf{x}_s$ into stego images $\mathbf{x}_{stego}$ , and transmits it through JSCC to the legitimate receiver. The receiver decrypts the stego images $\mathbf{x}’_{stego}$ using the same keys to obtain the secret images $\mathbf{x}’_{s}$.
\item \textbf{Key Agreement Center:} The key agreement center is responsible for private and public key storage and distribution. Different from traditional public key infrastructure (PKI), where the key pairs are almost fixed numbers, our key agreement center updates key pairs based on the transmission information, which is task-driven and content-dependent. The private key is determined by the content of secret image, while the public key is generated from the private key by modifying the core semantic meaning. The interaction between legitimate users and the key management center is assumed to be conducted through secure channels \cite{wang2025efficient}. 
\item \textbf{Physical Channel:} Additive white Gaussian noise (AWGN) is modeled as the physical channel \cite{xu2023latent}. The noise component $n$ follows a complex Gaussian distribution $\mathcal{CN}(0, \sigma_w^2)$, where $\sigma_w^2$ indicates the noise power. Hence, the stego images transmitted through channel as follows:
\begin{equation}
\label{0}
\mathbf{x}’_{stego} = h * \mathbf{x}_{stego} + n,
\end{equation}
where $h$ is the coefficient of the physical channel.
\item \textbf{Eavesdropper:} Eavesdropper attempts to intercept transmitted semantic feature $\mathcal{S}’_{Eve}$ from channel by
\begin{equation}
\label{0}
\mathcal{S}’_{Eve} = h * \mathcal{S} + n,
\end{equation}
where $\mathcal{S}$ is the original semantic feature.
Assumed to have access to semantic decoder $\mathcal{D}_{sem}(\cdot)$, eavesdropper tries to recover the transmitted image $\mathbf{x}’_{trans}$ by
\begin{equation}
\mathbf{x}’_{trans} =\mathcal{D}_{sem}(\mathcal{S}'_{Eve}).
\end{equation}
Based on the covertness of SemSteCom, the eavesdropper typically assumes that the decoded stego image $\mathbf{x}’_{trans}$ corresponds to the real secret image. To further assess the security of the proposed framework, we also consider a worst-case scenario that the eavesdropper is aware that steganography is employed and actively attempts to break the stego image by LLM or other advanced generative models, given by
\begin{equation}
\label{eq:eve_generative_attack}
\mathbf{x}’_{Eve}
= \tilde{\mathcal{G}}\!\left(
\mathbf{x}’_{trans};\, \tilde{\boldsymbol{\phi}}
\right),
\end{equation}
where $\mathbf{x}’_{Eve}$ denotes the secret image reconstructed by the eavesdropper, $\tilde{\mathcal{G}}(\cdot)$ represents a potential steganographic inversion function constructed by the eavesdropper, and $\tilde{\boldsymbol{\phi}}$ denotes the parameters of the reconstruction function.

\end{itemize}

\begin{table}[htbp]
\centering
\caption{The list of main parameters}
\begin{tabular}{ll}
\toprule
\textbf{Notation} & \textbf{Meaning} \\
\midrule
$\bar{\alpha}$ & Noise scheduling coefficient \\
$\mathcal{D}(\cdot)$ & Decoder \\
$\mathcal{E}(\cdot)$ & Encoder \\
$\mathcal{S}$ & Semantic latent representation\\
$\epsilon$ & Random Gaussian noise \\
$\epsilon_\theta$ & Predicted Gaussian noise \\
$E$ & Learnable projection matrix \\
$h$ & Public key token representation \\
$K_{priv}$ & Private semantic key \\
$K_{pub}$ & Public semantic key  \\
$L$ & Transformer layer number \\
$T$ & DDIM step \\
$t$ & Private key token representation \\
$W$ & Key projection matrix\\
${\mathbf{x}_s}$ & Secret image \\
${\mathbf{x}_{stego}}$ & Stego image \\
${z}$ & Latent variable during diffusion \\
\bottomrule
\end{tabular}
\label{tab:notations}
\end{table}

\subsection{Overview of the 
SemSteDiff Scheme}

The proposed SemSteDiff mainly comprises three modules, including the BLIP-based semantic key generation, conditional diffusion-based stego image generation, and JSCC-based semantic codec modules. The main parameters are illustrated in Table~\ref{tab:notations}. The steps of SemSteDiff are presented in Algorithm \ref{alg:semstediff}, with detailed descriptions as follows.

\subsubsection{BLIP-based Semantic Key Generation Module}

With the powerful cross-modal generation capability of BLIP\cite{li2022blip}, we propose a semantic key generation module to obtain keys by extracting textual descriptions from transmitted images. This design tightly links keys to image contents, achieving the idea of Semantics-as-Key. Formally, given a secret image $\mathbf{x}_s$, the BLIP model generates a semantic description as the private key as follows:
\begin{equation}
\label{1}
K_{\text{priv}} = f_\text{BLIP}(\mathbf{x}_s),
\end{equation}
where $K_{\text{priv}}= (K_1, K_2, \dots, K_T)$ is a sequence of semantic tokens. To enable secure coverless steganography, private key $K_{\text{priv}}$ is modified using a LLM to obtain a natural-language-style public key as
\begin{equation}
\label{2}
{K}_{\text{pub}} = f_\text{LLM}({K}_{\text{priv}}),
\end{equation}
which guides the diffusion pathway to generate the stego image.

\subsubsection{Conditional Diffusion Model-based Coverless Steganography Module}
To realize coverless stego image generation, we design a conditional diffusion-based module. This module directly generates stego images from latent representations, guided by semantic keys at both forward diffusion and reverse denoising stages.
\paragraph{Transmitter}
At the transmitter, the secret image is first encoded into a latent vector $\mathbf{z}_s$, followed by the forward diffusion process guided by $K_{\text{priv}}$, obtaining a noisy latent representation $\mathbf{z}_T$ using (\ref{3}). Subsequently, the public key \( K_{\text{pub}} \) is used to get the stego latent representation \( \hat{\mathbf{z}}_0 \) by guiding the reverse denoising process as (\ref{4})\cite{song2020denoising}. 
\begin{equation}
\label{3}
\mathbf{z}_T = \text{DDIM}(\mathbf{z}_s, \epsilon_\theta, K_{\text{priv}}, 0, T) \end{equation}
\begin{equation}
\label{4}
\hat{\mathbf{z}}_0 = \text{DDIM}(\mathbf{z}_T, \epsilon_\theta, K_{\text{pub}}, T, 0).
\end{equation}
Then, the VAE-based decoder converts latent domain \( \hat{\mathbf{z}}_0 \) into bit domain to reconstruct stego image \( \mathbf{x}_{\text{stego}} \).
\paragraph{Receiver}
At the receiver, we employ the same steps as the transmitter while exchanging the order of \( K_{\text{priv}} \) and  \( K_{\text{pub}} \) to achieve the reverse process by
\begin{equation}
\label{5}
\mathbf{z}_T' = \text{DDIM}(\mathbf{z}_{stego}; \epsilon_\theta, K_{\text{pub}}, 0, T)
\end{equation}
\begin{equation}
\label{6}
\hat{\mathbf{z}}'_0 = \text{DDIM}(\mathbf{z}'_T; \epsilon_\theta, K_{\text{pri}}, T, 0),
\end{equation}
where $\mathbf{z}_{stego}$ is the latent representation of transmitted stego image, $\mathbf{z}'_T$ is the recovery noisy latent representation, and $\hat{\mathbf{z}}'_0$ is the recovery latent representation of secret image. 

This module provides a steganographic mechanism that encodes secret image within the generative trajectory, instead of modifying existing cover images. It offers a  “invisible encryption” paradigm for high-security.
\subsubsection{JSCC-based Semantic Codec Module}
We consider an image SemCom over a wireless channel, where the goal is to transmit the stego images with minimal semantic distortion under noisy conditions. To this end, we utilize a JSCC-based semantic codec that directly learns a mapping between the visual domain and physical channel domain\cite{bourtsoulatze2019deep}. Formally, the stego image $\mathbf{x}_{\text{stego}}$ is encoded into a semantic feature $\mathcal{S}$ via a neural semantic encoder $\mathcal{E}_{sem}(\cdot)$ by
\begin{equation}
\label{7}
\mathcal{S} = \mathcal{E}_{sem}(\mathbf{x}_{\text{stego}}).
\end{equation}
This latent representation is directly transmitted over a noisy channel with AWGN, getting a disturbed signal $\mathcal{S}'$. At the receiver, a decoder $\mathcal{D}_{sem}(\cdot)$ reconstructs the image $\hat{\mathbf{x}}_{\text{stego}}$ by
\begin{equation}
\label{8}
\hat{\mathbf{x}}_{\text{stego}} =\mathcal{D}_{sem}(\mathcal{S}').
\end{equation}

\begin{algorithm}[t]
\caption{The Steps of the proposed SemSteDiff scheme}
\label{alg:semstediff}
\begin{algorithmic}[1]
\Statex \noindent\hspace*{-1.8em} \textbf{Stage 1: Private and Public Key Generation}
\State Generate private key ${K}_{\text{priv}}$ from secret image $\mathbf{x_s}$ by (\ref{1})

\State Produce public key ${K}_{\text{pub}}$ from ${K}_{\text{priv}}$ by (\ref{2})

\Statex \noindent\hspace*{-1.8em} \textbf{Stage 2: Stego Image Generation at Transmitter}
\State Encode secret image into latent space $\mathbf{z}_s$
\State Forward diffusion process under ${K}_{\text{priv}}$ guidance by (\ref{3})
\State Reverse denoising process under ${K}_{\text{pub}}$ guidance by (\ref{4})
\State Decode $\hat{\mathbf{z}}_0$ to get the stego image $\mathbf{x}_{stego}$

\Statex \noindent\hspace*{-1.8em} \textbf{Stage 3: JSCC-based Semantic Transmission}
\State Encode stego image using semantic encoder as (\ref{7})
\State Transmit $\mathcal{S}$ through wireless channel 
\State Decode transmitted $\mathcal{S}'$ using semantic decoder as (\ref{8})

\Statex \noindent\hspace*{-1.8em} \textbf{Stage 4: Secret Image Recovery from Receiver}
\State Encode $\mathbf{x}'_{stego}$ into latent space $\mathbf{z}_{stego}$
\State Forward diffusion process under ${K}_{\text{pub}}$ guidance by (\ref{5})
\State Reverse denoising process under ${K}_{\text{pri}}$ guidance by (\ref{6})
\State Decode $\hat{\mathbf{z}}'_0$ to get the reconstructed secret image $\mathbf{x}'_s$

\end{algorithmic}
\end{algorithm}
\section{Private and Public Key Generation}

\subsection{BLIP for Private Key Generation}

Motivated by \cite{yu2023cross} and \cite{yang2024diffstega}, we employ BLIP model as a private key generator to extract semantic features from images and use them as conditional inputs to guide the diffusion model in generating semantically consistent stego images. Compared to randomly generated digital keys, the natural language description itself as the key has obvious advantages. First, such keys are the actual content descriptions of images, enabling precise guidance of diffusion to stego images, which has strong stability. Second, since each image corresponds to a different and content-specific description, the generated keys are unique and random, thereby enhancing the security of the system. Additionally, BLIP specializes in matching image-text pairs and expressing complex scenes. It can avoid subjective biases in manual design while preventing the risk of key leakage caused by human errors.

\begin{figure*}[htpb]
    \centering
    \begin{subfigure}[b]{0.52\textwidth}
        \includegraphics[width=\linewidth]{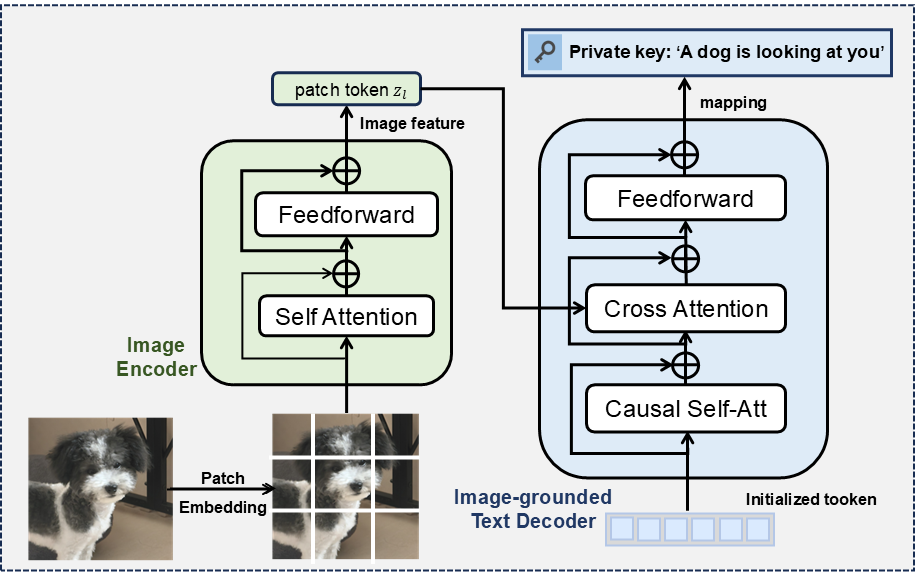}
        \caption{BLIP for Private Key Generation}
        \label{fig:blip}
    \end{subfigure}
    \hspace{1mm}  
    \begin{subfigure}[b]{0.44\textwidth}
        \includegraphics[width=\linewidth]{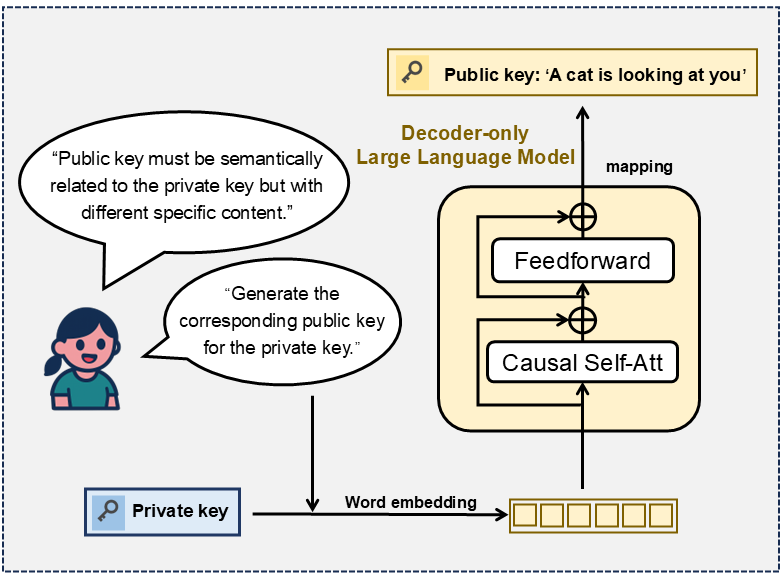}
        \caption{LLM for Public Key Generation}
        \label{fig:llm}
    \end{subfigure}

    \caption{Private and public key generation, where (a) shows that the private key is extracted from a secret image using BLIP model, (b) shows that public key is obtained by modifying private key under steganography requirement.}
    \label{figure:Blip}
\end{figure*}

The secret image $\mathbf{x_s} \in \mathbb{R}^{H \times W \times C}$ is first partitioned into $P \times P$ non-overlapping patches. Each patch is flattened and linearly projected into a fixed-dimensional embedding\cite{dosovitskiy2020image}. These patch tokens, together with a class token to capture global semantics $\mathbf{x}_{\text{class}}$, are concatenated to form the input sequence of image encoder, with positional embeddings added to retain spatial structure as
\begin{equation}
\label{9}
\mathbf{p}_0 = [\mathbf{x}_{\text{class}}; \mathbf{x_s}^1 \mathbf{E}; \ldots; \mathbf{x_s}^N \mathbf{E}] + \mathbf{E}_{\text{pos}},
\end{equation}
where $\mathbf{x_s}^i$ denotes the $i$-th image patch, $\mathbf{E}$ is the learnable projection matrix, and $\mathbf{E}_{\text{pos}}$ is the positional embedding.
The input $\mathbf{p}_0$ is passed through $L$ transformer encoder layers to extract semantic features. At each layer $\ell$, the intermediate state $\mathbf{p}'\ell$ is first computed via multi-head self-attention $\mathrm{MSA}(\cdot)$ over the normalized input $\mathrm{LN}(\cdot)$ from the previous layer by (\ref{10}). Subsequently, the hidden representation $\mathbf{p}\ell$ is obtained by applying a feedforward transformation $\mathrm{MLP}(\cdot)$ to $\mathbf{p}'_\ell$, as shown in (\ref{11}).

\begin{equation}
\label{10}
\mathbf{p}'_\ell = \mathrm{MSA}(\mathrm{LN}(\mathbf{p}_{\ell-1})) + \mathbf{p}_{\ell-1},
\end{equation}
\begin{equation}
\label{11}
\mathbf{p}_\ell = \mathrm{MLP}(\mathrm{LN}(\mathbf{p}'_\ell)) + \mathbf{p}'_\ell.
\end{equation}
After $L$ layers, the class token output from the final layer $\mathbf{p}_L$ is extracted as the global semantic representation of the secret image by
\begin{equation}
\label{12}
\mathbf{y} = \mathrm{LN}(\mathbf{p}_L^0).
\end{equation}

To generate a semantic private key, we use a image-ground text decoder that autoregressively predicts tokens \cite{li2022blip}. The decoder is initialized with a sequence of learnable tokens $\mathbf{t}_0$.
At each decoder layer $\ell$, the intermediate representation $\mathbf{t}'_\ell$ is computed using causal self-attention $\mathrm{CSA}(\cdot)$ by
\begin{equation}
\label{13}
\mathbf{t}'_\ell = \mathrm{CSA}(\mathrm{LN}(\mathbf{t}_{\ell-1})) + \mathbf{t}_{\ell-1}.
\end{equation}
To incorporate image context, we introduce a hidden state $\mathbf{h}\ell$ by applying cross-attention $\mathrm{CA}(\cdot)$ between the normalized $\mathbf{t}'{\ell}$ and the global image representation $\mathbf{y}$ as
\begin{equation}
\label{14}
\mathbf{h}_\ell = \mathrm{CA}(\mathrm{LN}(\mathbf{t}'_\ell), \mathbf{y}) + \mathbf{t}'_\ell.
\end{equation}
Then, the private token is obtained via a feedforward network by
\begin{equation}
\label{15}
\mathbf{t}_\ell = \mathrm{MLP}(\mathrm{LN}(\mathbf{h}_\ell)) + \mathbf{h}_\ell.
\end{equation}
Finally, the decoder output $\mathbf{t}_{L}^i$ is projected to the vocabulary probability distribution of private key $\hat{\mathbf{K}}_i$ via a softmax layer by
\begin{equation}
\label{16}
\hat{\mathbf{K}}_i = \mathrm{Softmax}(\mathbf{W}_{priv} \cdot \mathrm{LN}(\mathbf{t}_{L}^i)), 
\quad i = 1,\ldots, T,
\end{equation}
where $\mathbf{W}_{priv}$ is the private key projection matrix, and $i$ is the key position.
By sampling from $\hat{\mathbf{K}}_i$, we obtain the semantic private key as
\begin{equation}
\label{17}
K_{\text{priv}} = (K_1, K_2,\ldots, K_T).
\end{equation}

\begin{algorithm}[]
\caption{Steps of BLIP-based private key extractor}
\label{alg:PRIVATE_KEY}
\begin{algorithmic}[1]
\Statex \textbf{Input:} Secret images

\State Divide secret image $\mathbf{x_s}$ into non-overlapping patches and encode them as visual tokens with positional embeddings by (\ref{9});
\State Extract high-level semantic features through self-attention and multilayer perceptron (MLP) layers from (\ref{10}) to (\ref{11});
\State Obtain global secret image representation $\mathbf{y}$  by (\ref{12});
\State Initialize decoder tokens to begin private key generation;
\State Construct private token dependencies via causal self-attention by (\ref{13});
\State Inject image semantics into token generation via cross-attention with $\mathbf{y}$ as (\ref{14});
\State Enhance private token features with feedforward network by (\ref{15});
\State Predict token distributions using softmax over vocabulary space by (\ref{16});
\State Decode final private key $K_{\text{priv}} = (K_1, K_2,\ldots, K_T)$ through sequential sampling by (\ref{17});
\Statex \textbf{Output:} Private keys
\end{algorithmic}
\end{algorithm}

\begin{rem}
As illustrated in Algorithm \ref{alg:PRIVATE_KEY}, the private key generation scheme encapsulates the semantic representation of secret images and functions as a cross-modal embedding for secure SemCom. The semantic key $K_{\text{priv}}$ is adaptively derived from the content of each individual image. This specific encoding ensures that private keys remain semantically aligned with the various visual inputs. 
\end{rem}

\subsection{LLM for Public Key Generation}
The purpose of a coverless steganography task is to generate stego images that do not arouse suspicion of eavesdroppers. Under a specific communication background, eavesdroppers may have a general expectation of the transmitted contents, which leads to a certain restriction of public keys. Due to this situation, the public key must be semantically related to the private key but with different specific content, ensuring the generative diffusion scene guided by public key has similar distributions with secret image, yet remaining content-independent. Even if an eavesdropper knows the class of the transmitted content, the generated stego image will not reveal the true secret information. Based on high requirements for the public keys, we introduce LLMs to generate public key prompts due to their stronger contextual understanding, linguistic diversity, and control capability\cite{minaee2024large}. In the previous work \cite{cao2024multimodal}, the LLM is used for converting images to a textual descriptions, then the texts are encrypted and transmitted to the receiver, and finally the receiver decrypts the texts and recovers the original images based on the content. In contrast, we serve the LLM-generated prompts as public keys directly, without additional encryption. Specific LLM-required prompt is designed to control the generation process only preserving general category and background of private key sentence, while detailed descriptions related to objects, attributes, or behavior are intentionally replaced
with irrelevant counterparts, which ensures the public key remains semantic relative to private key without any leakage. This semantics-as-key design simplifies the communication process while still preserving the security of the hidden information.

\begin{algorithm}[t]
\caption{Steps of LLM-based public key generator}
\label{alg:PUBLIC_KEY}
\begin{algorithmic}[1]
\Statex \textbf{Input:} Private key $K_{\text{priv}}$
\State Initialize private key tokens for autoregressive decoding;
\State Compute public token representations via causal self-attention and MLP as (\ref{18})–(\ref{19});
\State At each decoding step, obtain public token probability distribution using softmax over vocabulary as (\ref{20});
\State Sample predicted token $K_i'$ from the distribution to form the final public key sequence $K_{\text{pub}} = (K_1', K_2',\ldots, K_T')$ by (\ref{21});
\Statex \textbf{Output:} Public key $K_{\text{pub}}$
\end{algorithmic}
\end{algorithm}

We use the generated private key sequence $K_{\text{priv}}$ as the initial prompt input to a decoder-only LLM to generate the public key. At each layer $\ell$, the public token representation $\mathbf{q}_\ell$ is updated by


\begin{equation}
\begin{aligned}
\label{18}
\mathbf{q}'_\ell &= \mathrm{CSA}(\mathrm{LN}(\mathbf{q}_{\ell-1})) + \mathbf{q}_{\ell-1} 
\end{aligned}
\end{equation}
and
\begin{equation}
\begin{aligned}
\label{19}
\mathbf{q}_\ell  &= \mathrm{MLP}(\mathrm{LN}(\mathbf{q}'_\ell)) + \mathbf{q}'_\ell.
\end{aligned}
\end{equation}

At each decoding step, the public key token distribution $\hat{\mathbf{K}}'_i$ is computed as
\begin{equation}
\label{20}
\hat{\mathbf{K}}'_i = \mathrm{Softmax}(\mathbf{W}_{pub} \cdot \mathrm{LN}(\mathbf{q}_L^i)), \quad i = 1, \ldots, T,
\end{equation}
where $\mathbf{W}_{pub}$ is the private key projection matrix.
Same as the private key generation step. The predicted token $K'_i$  is sampled from \( \hat{\mathbf{K}}'_i \) to form the final public key sequence $K_{\text{pub}}$ as

\begin{equation}
\label{21}
K_{\text{pub}} = (K'_1, K'_2, \dots, K'_T).
\end{equation}

To ensure that the public key remains semantically related to the private key while remaining sufficiently distinct to mislead eavesdroppers, several public keys towards on one private key are generated to avoid generation collapsing. The semantic similarity between public key-guided stego images and secret images, as well as the correlation between public keys and stego images are both considered by evaluating embedding-based similarity such as Contrastive Language–Image Pre-training (CLIP) score\cite{radford2021learning} to comprehensively to choose the optimal one. It jointly prevent semantic leakage caused by over-specific descriptions and avoid uninformative public keys, which further reduces the likelihood of generating similar public keys across different secret images, improving the robustness and security of the LLM-based key generation process in SemSteDiff.

\begin{rem}
By utilizing LLM's generation diversity, our scheme can finely tune private key prompts to get public key shown in Algorithm \ref{alg:PUBLIC_KEY}, leading to a generative diffusion image that has a similar distributions with secret images but remain content-independent, which can further enhance the semantic covertness of the stego image.
\end{rem}


\section{Conditional Diffusion Model-based Coverless Steganography Module}

\begin{figure*}[t]
\centering
\vspace{-10mm}
\includegraphics[width=0.95\linewidth]{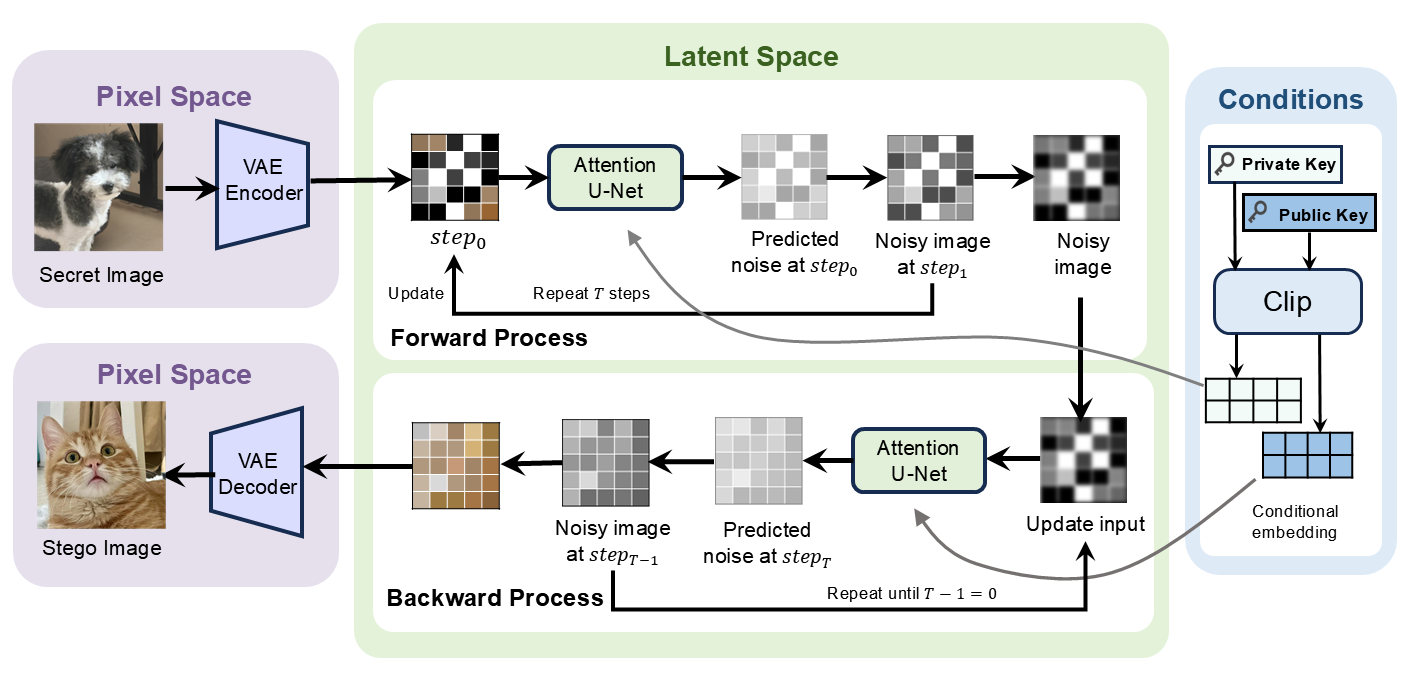}
\caption{Process of conditional diffusion model-based coverless steganography. First, the secret image encodes from pixel space into latent space by VAE encoder. Second, the secret latent vector adds noise under the guidance of private key using attention mechanism. Then, the public key guides the noisy vector to generate the stego latent vector. Finally, the VAE decoder decodes the latent vector into stego image.}
\label{fig411}
\end{figure*}

The proposed conditional diffusion model-based coverless steganography module is illustrated as Figure \ref{fig411}. This module includes two parts: the Latent Steganography Framework and the Key-Guided Stego Image Generation.

\subsection{VAE-based Latent Steganography Framework}

In traditional image steganography, secret image is typically embedded directly into the pixel space of the cover image\cite{yu2023cross}. However, such approach is often susceptible to compression and noise , making it difficult to balance accuracy and security. To address this limitation, we covert the steganographic generation process from the observable image domain to compact latent space by introducing a VAE-based latent modeling strategy\cite{rombach2022high}.

\begin{algorithm}[t]
\caption{VAE-based Latent Modeling for Stego Image}
\label{alg:vae-combined}
\begin{algorithmic}[1]
\Statex \hspace{-\algorithmicindent} \textbf{Training Phase:}
\Repeat
    \State Sample images $\mathbf{x} \sim q(\mathbf{x})$ \
    \State Extract the mean and standard deviation of the normal distribution $(\mu, \log \sigma^2) = \mathcal{E}_\text{VAE}(\mathbf{x})$ 
    \State Add disturbance $\epsilon \sim \mathcal{N}(0, \mathbf{I})$ 
    \State Reparameterize latent vector $\mathbf{z} = \mu + \sigma \odot \epsilon$ 
    \State Reconstruct the image $\hat{\mathbf{x}} = \mathcal{D}_\text{VAE}(\mathbf{z})$ 
    \State Compute loss including reconstruction term and regularization term by (\ref{22})
    \State Update parameters $\phi$ and $\theta$ by gradient descent
\Until{converged}
\vspace{0.3em}
\Statex \hspace{-\algorithmicindent} \textbf{Inference Phase:}
\Statex \textbf{Input:} secret image $\mathbf{x}_s$
\State Encode the secret image by (\ref{23})
\State Sampling standard Gaussian perturbation $\epsilon \sim \mathcal{N}(0, \mathbf{I})$
\State Compute latent vector $\mathbf{z}_s$ by (\ref{24})
\State Recover the latent vector into pixel space by (\ref{25})
\State \Return reconstructed secret image $\mathbf{x}'_s$
\end{algorithmic}
\end{algorithm}

During the training phase, the encoder maps the image into a Gaussian posterior characterized by mean $\mu$ and variance $\sigma^2$, from which the latent code is sampled via the reparameterization trick. The decoder reconstructs the image from this latent code to calculate the reconstruction loss as the first term in (\ref{22}) , while a Kullback–Leibler(KL) divergence regularizes the latent distribution as the second term in (\ref{22}) by 
\begin{equation}
\begin{aligned}
\label{22}
\mathcal{L} = \|\mathbf{x} - \hat{\mathbf{x}}\|^2 + D_{\mathrm{KL}} \big( \mathcal{N}(\mu, \sigma^2) \, \| \, \mathcal{N}(0, \mathbf{I}) \big),
\end{aligned}
\end{equation}
where $\mathbf{x}$ is the original image, $\hat{\mathbf{x}}$ is the reconstructed image, $D_{\mathrm{KL}}$ measures the KL divergence between the Gaussian posterior and the standard normal prior.

During the inference phase, the secret image $\mathbf{x}_s$ is first passed through a variational encoder $\mathcal{E}_\text{VAE}(\cdot)$ to obtain the latent distribution by
\begin{equation}
\label{23}
(\mu, \log \sigma^2) = \mathcal{E}_\text{VAE}(\mathbf{x}_s).
\end{equation}
Then, we perform stochastic sampling from this latent distribution by applying the reparameterization trick as below:

\begin{equation}
\begin{aligned}
\label{24}
\mathbf{z}_s = \mu + \sigma \odot \epsilon,
\end{aligned}
\end{equation}
where $\epsilon$ is a standard Gaussian noise vector. The sampled latent variable $\mathbf{z}_s$ serves as the starting point for the following diffusion process. After steganography and semantic transmission, the variational decoder $\mathcal{D}_\text{VAE}(\cdot)$ transforms the latent variable $\mathbf{z}'_s$ back into the bit space to reconstruct the secret image by
\begin{equation}
\label{25}
\mathbf{x}'_s = \mathcal{D}_\text{VAE}(\mathbf{z}'_s),
\end{equation}
where $\mathbf{x}'_s$ denotes the reconstructed secret image.
\begin{rem}
As illustrated in Algorithm \ref{alg:vae-combined},  we employ a VAE\cite{kingma2013auto} to encode an input image $\mathbf{x}$ into a latent representation $\mathbf{z}$ that captures its semantic feature. The steganographic capacity is constrained by the VAE-based latent space $\mathcal{Z} \in \mathbb{R}^{h \times w \times c}$, where each secret image is mapped into a compact latent vector. The capacity per generated stego image is simply calculated as $C_L = (h \times w \times c) \times b$ bits, where $b$ represents the numerical precision. This latent representation retains the semantic core of the image and suppresses redundant pixel-level details to offer a more compact encoding, which also provides a stable and controllable foundation for the subsequent key-guided diffusion generation.

\end{rem}

\subsection{Key-Guided Stego Image Generation}
To effectively integrate the key prompt into the diffusion process, we do not simply concatenate it with image features but instead using an attention mechanism for semantic alignment. The attention mechanism aims to determine “where to pay attention” and “what to pay attention”, thereby establishing a precise correspondence between image contents and key conditions\cite{vaswani2017attention}.

In our framework, the U-Net architecture serves as the core denoising network of the latent diffusion model. It takes the latent variable $\mathbf{z}_t$ and timestep $t$ as the input to predict the noise component. Owing to its encoder-decoder design, the U-Net effectively captures both local and global features, making it well-suited for image reconstruction in latent space. To further enhance the conditional guidance, attention mechanism is embedded into the intermediate layers of U-Net to integrate keys into diffusion process. 

The generated keys are embedded into latent space by CLIP model, which is a multimodal neural network to extract textual embedding \cite{radford2021learning},given as follows.
\begin{equation}
\label{25.5}
E_c = f_{\text{CLIP}}(\text{Key}).
\end{equation}
The embedding key prompt $E_c$ is projected to obtain keys $K$ and values $V$ and the sampled latent variable $\mathbf{z}_t$ is used to calculate query $Q$ by
\begin{equation}
\label{26}
Q = W_Q \mathbf{z}_t, \quad K = W_K E_c, \quad V = W_V E_c,
\end{equation}
where $W_Q$, $W_K$ and $W_V$ are trainable parameters, $t$ is the sampling step \cite{vaswani2017attention, jaegle2021perceiver}. It combines semantic keys with diffusion process by controling every sampling steps.
The semantic feature correlation is computed by
\begin{equation}
\label{27}
\text{CrossAttn}(\mathbf{z}_t, E_c) = \text{softmax}\left(\frac{QK^\top}{\sqrt{d}}\right) V,
\end{equation}
where $d$ is scaling factor. By calculating the correlation between the query and the key, we obtain a set of attention weights, which determine which parts of the semantic vectors should be referenced by different positions in the image during the generation process\cite{rombach2022high, reed2016generative}. 

This attention design enables each denoising step to attend to the key-conditioned semantics dynamically. As shown in Figure \ref{fig411}, the key embeddings, whether derived from a private key or public key, flow into the latent-space denoising module via attention-aware mechanism. Finally, the fused attention outputs are injected into the feature path of the U-Net, guiding the model to reconstruct the image consistent with the semantic key.

In addition, we adopt classifier-free guidance\cite{ho2022classifier} to enhance the flexibility of semantic control. The attention output is fused into the latent feature and passed through a noise prediction model $f_\theta$. The model predicts conditional noise $\hat{\varepsilon}_{\text{cond}}$ and unconditional noise $\hat{\varepsilon}_{\text{uncond}}$ by (\ref{28}) and (\ref{29}) separately.
\begin{equation}
\label{28}
\hat{\varepsilon}_{\text{cond}}(\mathbf{z}_t, t, Key) = f_\theta(\mathbf{z}_t, t) + \text{CrossAttn}(\mathbf{z}_t, E_c),
\end{equation}
\begin{equation}
\label{29}
\hat{\varepsilon}_{\text{uncond}}(\mathbf{z}_t, t) = f_\theta(\mathbf{z}_t, t) + \text{CrossAttn}(\mathbf{z}_t, E_\emptyset),
\end{equation}
where $E_\emptyset$ is the null embedding vector, and \textit{Key} is applied to compute the $E_c$  in (\ref{25.5}).
Then, conditional and unconditional noise are weighted using a guidance scale factor $\beta$ to balance semantic consistency and image realism:
\begin{equation}
\label{30}
\hat{\epsilon}_{\text{final}} = \hat{\epsilon}_{\text{uncond}} + \beta (\hat{\epsilon}_{\text{cond}} - \hat{\epsilon}_{\text{uncond}}).
\end{equation}
A larger guidance factor causes the generated images to align more strictly with the semantic contents of the key prompts, while a smaller factor prioritizes the naturalness of the images\cite{ho2022classifier}. Through this attention-guided conditional diffusion modeling approach, semantic keys can be precisely and covertly embedded into the image generation process.

The dual-conditioned generation process incorporates both a private key and a public key to jointly control the stego image generation, which is divided into two symmetric diffusion stages.
\begin{enumerate}
\item 
\textbf{Private-Key Reverse Denoising to Synthesize Noise:}
The generation process begins with a reverse diffusion procedure driven by the private key $K_{\text{priv}}$. Instead of transmitting the image directly, the $K_{\text{priv}}$ guides a forward deterministic diffusion process from latent $\mathbf{z}_0$ to $\mathbf{z}_T$ over $T$ steps \cite{song2020denoising}.
This inversion process follows the forward DDIM update by

\begin{equation}
\mathbf{z}_{t+1} = \sqrt{\bar{\alpha}_{t+1}} \left( \frac{\mathbf{z}_t - \sqrt{1 - \bar{\alpha}_t} \cdot \hat{\epsilon}_t}{\sqrt{\bar{\alpha}_t}} \right) + \sqrt{1 - \bar{\alpha}_{t+1}} \cdot \hat{\epsilon}_{t, \text{priv}},
\label{31}
\end{equation}
where $\hat{\epsilon}_{t, \text{priv}}$ is the semantic-conditioned noise prediction under private key guidance, and $\bar{\alpha}$ is the noise scheduling coefficient. In (\ref{30}), $\sigma_t = 0$ makes the transformation fully deterministic. The output $\mathbf{z}_T$ serves as an encrypted latent embedding of the secret image, recoverable only through the correct key-driven inversion.
\item 
\textbf{Stego Sampling via Public Key:}
The public key $K_{\text{pub}}$ is then used to guide the reverse sampling from $\mathbf{z}_T$ back to a visually coherent stego latent $\hat{\mathbf{z}}_0$. The reverse DDIM update is defined as:

\begin{equation}
\mathbf{z}_{t-1} = \sqrt{\bar{\alpha}_{t-1}} \left( \frac{\mathbf{z}_t - \sqrt{1 - \bar{\alpha}_t} \cdot \hat{\epsilon}_t}{\sqrt{\bar{\alpha}_t}} \right) + \sqrt{1 - \bar{\alpha}_{t-1}} \cdot \hat{\epsilon}_{t, \text{pub}}.
\label{32}
\end{equation}
This ensures that the final output appears to follow the public key semantics, while the underlying generation path remains uniquely influenced by the private key trajectory.
\end{enumerate}

\setlength{\abovedisplayskip}{2pt}
\setlength{\belowdisplayskip}{2pt}

\begin{algorithm}[t]
\caption{The Key-Guided Conditional Diffusion Algorithm based on Attention Mechanism}
\label{alg:attention_diffusion}
\begin{algorithmic}[1]
\Statex \textbf{Input:} key prompt $c$, guidance scale $\beta$
\State Initialize noise sample image $\mathbf{x}_T \sim \mathcal{N}(0, I)$

\For{$t = T, T{-}1,\ldots, 1$}
    \State Sampling random disturbance:
    \Statex \centerline{$z \sim \mathcal{N}(0, I)$ if $t > 1$, else $z = 0$}
    \vspace{-1em}
    \State Encode key prompt as conditional embedding by (\ref{25.5})
    \State Project attention vectors by (\ref{26})
    \State Integrate key into image representation by (\ref{27})
    \State \parbox[t]{\dimexpr\linewidth-\algorithmicindent}{%
    Compute conditional and unconditional noise predictions using trained noise predictor $f_\theta$ by (\ref{28}) and (\ref{29})%
    }
    \vspace{0.5em}
    \State Compute guided noise by by (\ref{30})
    \State Update latent using DDIM step:
    \begin{equation*}
    \begin{aligned}
    \mathbf{z}_{t-1} =\ & \sqrt{\bar{\alpha}_{t-1}} \left( \frac{\mathbf{z}_t - \sqrt{1 - \bar{\alpha}_t} \cdot \hat{\epsilon}_{\text{final}}}{\sqrt{\bar{\alpha}_t}} \right) \\
    & + \sqrt{1 - \bar{\alpha}_{t-1} - \sigma_t^2} \cdot \hat{\epsilon}_{\text{final}} + \sigma_t z
    \end{aligned}
    \end{equation*}

\EndFor
\State \Return $\mathbf{z}_0$
\end{algorithmic}
\end{algorithm}

\begin{rem}
Unlike traditional explicit embedding steganography methods, our approach does not modify existing images but instead of performing steganography directly during image generation. As illustrated in Algorithm~\ref{alg:attention_diffusion}, by conditional sampling process based on semantic keys, the secret information is hidden into the generative path while the public information controls the content of stego images. 
\end{rem}

\section{Simulation Results and Analysis}

\subsection{Simulation Parameters}
\subsubsection{Datasets}
We utilize Stego260 datasets\cite{yu2023cross} to verify the performance of SemSteDiff. Stego260 consists of 260 natural images collected from two public datasets\footnote{\url{https://github.com/aisegmentcn/matting_human_datasets}}\footnote{\url{https://www.kaggle.com/datasets/iamsouravbanerjee/animal-image-dataset-90-different-animals}} and web sources. We also utilize the UniStega \cite{yang2024diffstega} as supplementary dataset to verify generalizability of our scheme, which are collected from multiple public datasets, including COCO, AFHQ, and CelebA-HQ, as well as other online sources. The images are resized as 512*512.

\subsubsection{Parameter Settings}

In SemSteDiff, the key generation module and coverless
steganography module are pre-trained. For the key generation  module, private key generation is employed BLIP-1 base model\footnote{\url{https://huggingface.co/Salesforce/blip-image-captioning-base}} to extract descriptions from secret images, and public key generation is employed application programming interface of ChatGPT 4o to dynamically generate public keys. The prompt is set as ``Given the following semantic description extracted from an image, generate a public description that preserves only category-level or scene-level semantics, and replaces instance-specific attributes such as appearance details, spatial relations, and identities with another different attribute" for ChatGPT to generate public keys. For the coverless steganography module, we use the Stable Diffusion v1.5\footnote{\url{https://huggingface.co/stable-diffusion-v1-5/stable-diffusion-v1-5}} in the forward and reverse diffusion process, with the 50 DDIM steps.
\subsubsection{Validation Framework}
The SemSteDiff is verified under the AWGN channel ranged from 0 dB to 10 dB SNR, and three different JSCC frameworks including DeepJSCC\cite{bourtsoulatze2019deep}, NTSCC\cite{dai2022nonlinear}, and SwinJSCC\cite{yang2024swinjscc} as the semantic codec module. For DeepJSCC and SwinJSCC, the compression ratio is fixed at $1/12$, while for NTSCC the maximum code rate $R$ is constrained to $1/16$ to enable a fair comparison across schemes\cite{dai2022nonlinear}. Each is trained under Signal-to-Noise Ratio (SNR) values of $0$, $2$, $4$, $6$, $8$, and $10$ dB, respectively. The training of JSCC models and test of SemSteDiff is performed on NVIDIA RTX~6000 Ada Generation GPU under PyTorch~2.7.1 with CUDA~11.8 enabled.

\subsection{Complexity Analysis}

We analyze the computational complexity of SemSteDiff based on models used in the simulation. The theoretical complexity focuses on three main parts: the BLIP-1 base model for private key extraction, the LLM for public key generation, the Stable Diffusion v1.5 for coverless image steganography. The overall complexity is determined by the total employment of these modules. We also measure the overhead consumption of SemSteDiff including inference latency, peak GPU memory consumption, and floating-point operations (FLOPs) under batch size of 1 and input resolution of 512×512 on a NVIDIA RTX 6000 Ada Generation GPU. The results are shown in Table \ref{tab:profiling}.

\begin{table}[t]
\centering
\caption{Overhead consumption of major modules in SemSteDiff, where FLOPs of latent diffusion are calculated per sampling step. As the overhead of ChatGPT is inaccessible, we choose lightweight LLM of ChatGLM \cite{glm2024chatglm} to measure the Overhead deployment of generating public key.}
\label{tab:profiling}
\fontsize{7pt}{8.5pt}\selectfont 
\setlength{\tabcolsep}{3pt}
\renewcommand{\arraystretch}{1.05}
\begin{tabular}{l c c c c}
\toprule
\textbf{Module} & \textbf{FLOPs} & \textbf{Precision} & \textbf{Latency (s)} & \textbf{Memory (MB)} \\
\midrule

\multirow{2}{*}{Private key extraction module} 
 & \multirow{2}{*}{129.28G} & FP32 & 0.10 & 921.02 \\
 &  & FP16 & 0.09 & 476.05 \\

\midrule
\multirow{2}{*}{Semantic codec module} 
 & \multirow{2}{*}{4.95G} & FP32 & 0.0017 & 25.25 \\
 &  & FP16 & 0.0012 & 22.17 \\

\midrule
\multirow{2}{*}{Coverless steganography module} 
 & \multirow{2}{*}{677.76G} & FP32 & 9.98 & 6227.86 \\
 &  & FP16 & 3.88 & 3258.50 \\

\midrule
Public key generation module
 & -- & -- & 0.96 & 11968.41 \\

\bottomrule
\end{tabular}
\end{table}

\subsubsection{Private key extraction} 
The BLIP captioning model comprises a ViT-based vision encoder and a transformer-based language decoder. Given an input image of size $H \times W$, the encoder extracts $N$ patches with patch size $P \times P$, and then processes them via $L$ transformer layers, each with embedding dimension $d$. The encoder complexity is $\mathcal{O_\text{BLIPEncoder}}(L \cdot N^2 \cdot d)$. The decoder, which generates a semantic key with maximum token length $T$ and $L'$ layers, has complexity determined by $\mathcal{O_\text{BLIPDecoder}}(L' \cdot T^2 \cdot d).$
\subsubsection{Public key generation} 
The LLM is responsible for the public key generation. In the simulation, we use application programming interface of ChatGPT to achieve high quality public key response. However, due to the inaccessibility of specific internal model architecture, the exact computational complexity of ChatGPT remains difficult to estimate. In this part, we model the LLM-based public key generation as an inference process of a decoder-only Transformer with key–value caching enabled. The private key is provided as an input prompt consisting of $T_{\text{in}}$ tokens, and the LLM autoregressively generates a public key with $T_{\text{pub}}$ tokens. The LLM is composed of $L_m$ Transformer layers with hidden dimension $d_m$. At each decoding step, the dominant computation arises from the linear projections and feed-forward operations within each Transformer layer, as well as the attention interaction between the newly generated token and the cached representations of the input private key. As a result, the overall computational complexity of public key generation is approximated as $\mathcal{O}_{\text{LLM}} \approx \mathcal{O}(L_m \cdot T_{\text{pub}} \cdot d_m^2) + \mathcal{O}(L_m \cdot T_{\text{pub}} \cdot T_{\text{in}} \cdot d_m)$. It is noted that the complex LLM-based key generation can be executed offline stage to decrease the computational cost in practical deployment.

\subsubsection{Coverless image steganography} 
The steganography is performed by Stable Diffusion v1.5, which includes a U-Net denoiser, a VAE, and a text encoder. The primary computational cost arises from the denoising process using DDIM with $T$ steps. In each step, the U-Net processes a latent tensor of size $C \times H \times W$, resulting in complexity as $\mathcal{O_\text{Diffusion}}(T \cdot C^2 \cdot H \cdot W)$. The VAE encoder and decoder only run once in each latent diffusion process, which contributes negligible overhead compared to iterative denoising.

Therefore, the overall complexity of the system is $\mathcal{O}(L \cdot N^2 \cdot d + L' \cdot T^2 \cdot d + T \cdot C^2 \cdot H \cdot W)+\mathcal{O}_{\text{LLM}}$. It is noted that the key generation is not involved in the diffusion sampling and JSCC processes. Instead, it can be advanced employed before transmission to prepare the relative keys separately, which means the BLIP and LLM could operate only once in practical deployment, significantly reducing computational cost during real-time transmission. Therefore, its computational cost does not scale with channel conditions, coding rate, or diffusion steps. Although additional overhead arises from interactions between modules, such overhead does not dominate the overall computational efficiency of the proposed framework.

\subsection{Performance Metrics}
To comprehensively evaluate the performance of SemSteDiff, four widely adopted image quality assessment metrics are employed, including peak signal-to-noise ratio (PSNR), structural similarity index measure (SSIM), learned perceptual image patch similarity (LPIPS), and mean squared error (MSE).

\textbf{1) PSNR:}  
PSNR quantifies the ratio between the maximum possible signal power and noise power:
\begin{equation}
\text{PSNR} = 10 \log_{10} \left( \frac{MAX_I^2}{\text{MSE}} \right),
\end{equation}
where $MAX_I$ is the maximum possible pixel value, and MSE is the mean squared error between the reconstructed and original images.

\textbf{2) SSIM:}  
SSIM measures the structural similarity between two images by considering luminance, contrast, and structural components\cite{hore2010image}. It is computed on local image patches and then averaged across the image by
\begin{equation}
\text{SSIM}(x, y) = \frac{(2\hat\mu_x\hat\mu_y + C_1)(2\hat\sigma_{xy} + C_2)}
{(\hat\mu_x^2 + \hat\mu_y^2 + C_1)(\hat\sigma_x^2 + \hat\sigma_y^2 + C_2)},
\end{equation}
where $x$ and $y$ are sampling batches of original image and reconstructed image, $\hat{\mu}$ and $\hat{\sigma}$ are mean value and variance of image patch, respectively, and $C$ is the stability constant. SSIM values range from $0$ to $1$, with higher values indicating greater structural similarity.

\textbf{3) LPIPS:}  
LPIPS is a perceptual similarity metric that computes the distance between deep feature representations extracted from a pretrained network  $\hat{\phi}_l(\cdot)$\cite{zhang2018unreasonable}. It can be measured by:
\begin{equation}
\text{LPIPS}(\mathbf{x}, \hat{\mathbf{x}}) = \sum_{l} w_l \cdot \frac{1}{H_l W_l} \sum_{h=1}^{H_l} \sum_{w=1}^{W_l} \left\| \hat{\phi}_l(\mathbf{x})_{hw} - \hat{\phi}_l(\mathbf{x}')_{hw} \right\|_2^2,
\end{equation}
where $H$ and $W$ represent height and width of the input images, respectively, and $w_l$ is the layer weight. Lower LPIPS scores indicate that the reconstructed image is perceptually closer to the reference.

\textbf{4) MSE:}  
MSE evaluates the average squared difference between the pixel intensities of the reconstructed and reference images:
\begin{equation}
\text{MSE} = \frac{1}{HW} \sum_{i=1}^H \sum_{j=1}^W \left[ I(i,j) - \hat{I}(i,j) \right]^2,
\end{equation}
where $I(i,j)$ is the the pixel intensities at position $(i,j)$. A smaller MSE indicates more accurate pixel-level reconstruction.

\textbf{5) CLIP Score:}
To further evaluate the semantic security of SemSteDiff, we also adopt the CLIP score \cite{radford2021learning}  to measure the semantic consistency between reconstructed images and private keys, which reflects the ability of legitimate receivers and eavesdroppers to recover secret information.
CLIP score measures the alignment between an image and a textual semantic description in a shared embedding space, providing a direct indicator of whether meaningful semantic information can be recovered. It measures the cosine similarity between reconstructed image $\hat{\mathbf{x}}$ and private key $K_{priv}$
\begin{equation}
\mathrm{CLIP}(K_{priv}, \hat{\mathbf{x}}) 
= \cos\left(\mathbf{v}_{priv}, \mathbf{v}_{img}\right)
= \frac{\mathbf{v}_{priv}^\top \mathbf{v}_{img}}{\|\mathbf{v}_{priv}\|_2 \|\mathbf{v}_{img}\|_2},
\end{equation}
where $\mathbf{v}_{img}$ and $\mathbf{v}_{priv}$ are normalized embedding vectors of reconstructed image $\hat{\mathbf{x}}$ and private key $K_{priv}$, respectively. A higher CLIP score indicates stronger semantic alignment between the reconstructed image and the target semantic description.



\begin{figure*}[]
    \centering
    \begin{subfigure}[b]{0.24\textwidth}
        \includegraphics[width=\linewidth]{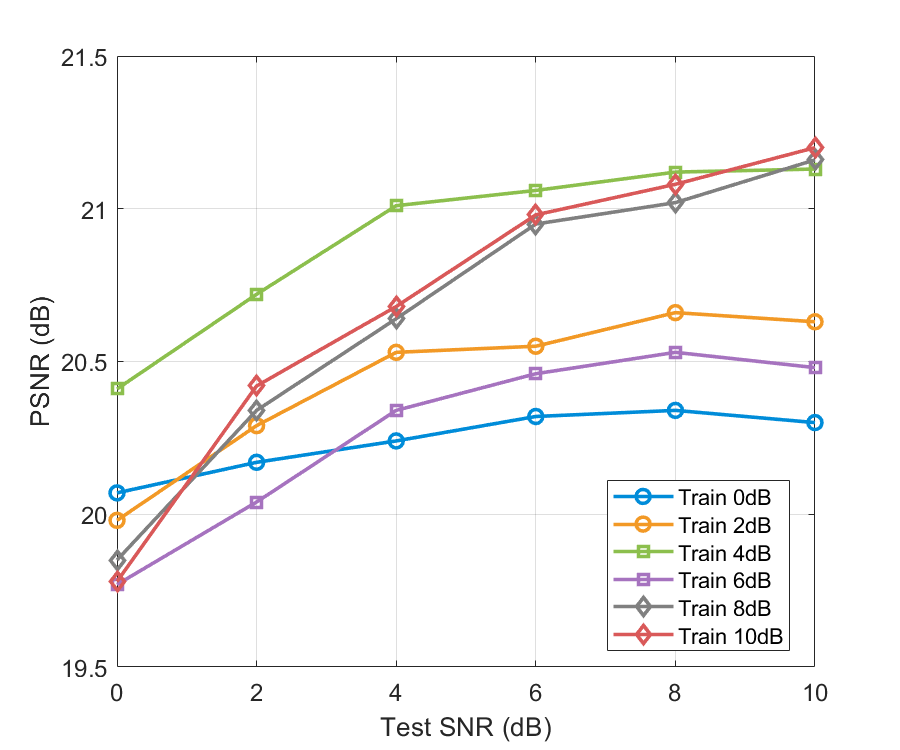}
        \caption{PSNR of SemSteDiff with DeepJSCC}
        \label{fig:deep+psnr}
    \end{subfigure}
    \hfill
    \begin{subfigure}[b]{0.24\textwidth}
        \includegraphics[width=\linewidth]{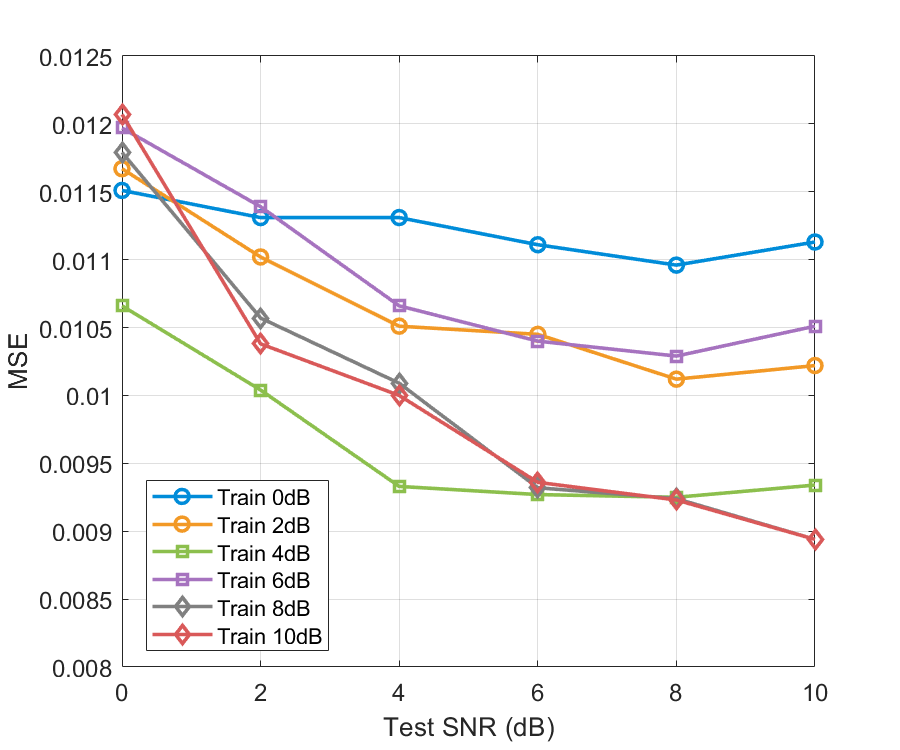}
        \caption{MSE of SemSteDiff with DeepJSCC}
        \label{fig:deep+mse}
    \end{subfigure}
    \hfill
    \begin{subfigure}[b]{0.24\textwidth}
        \includegraphics[width=\linewidth]{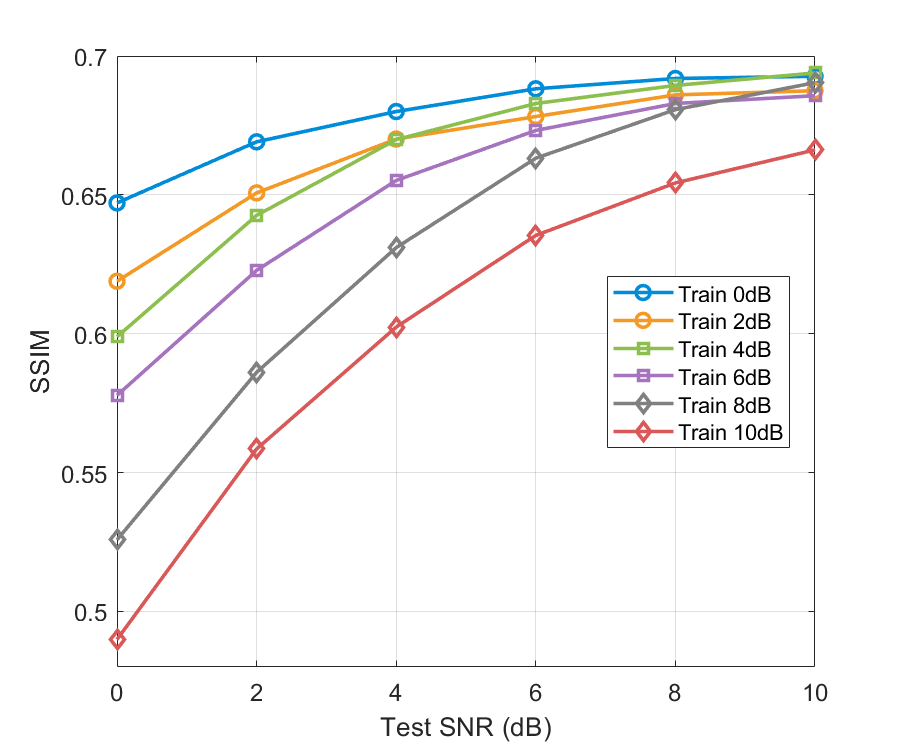}
        \caption{SSIM of SemSteDiff with DeepJSCC.}
        \label{fig:deep+ssim}
    \end{subfigure}
    \hfill
    \begin{subfigure}[b]{0.24\textwidth}
        \includegraphics[width=\linewidth]{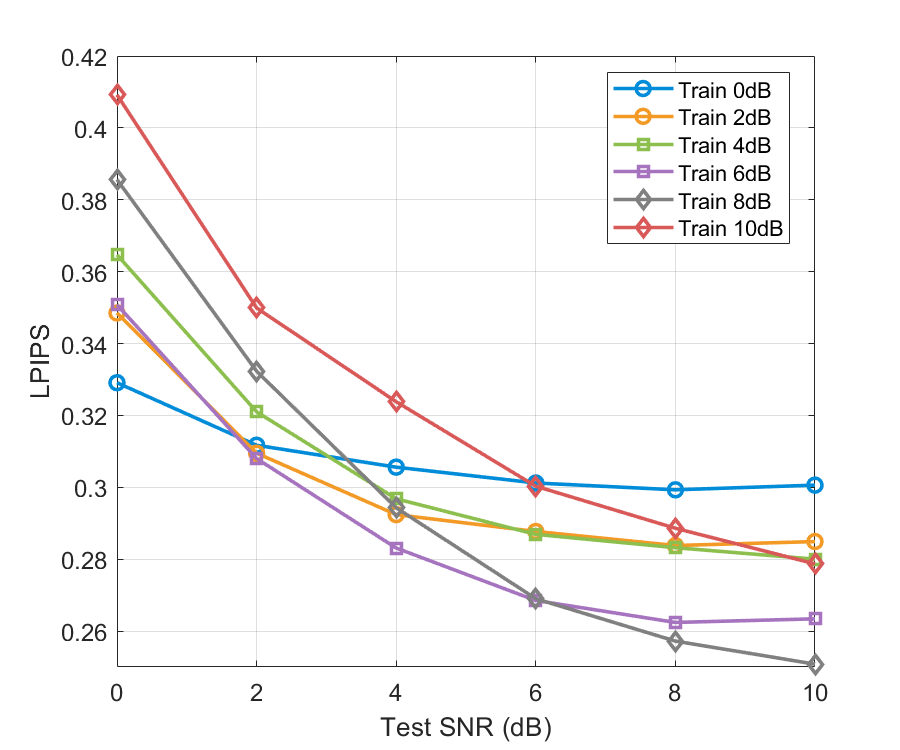}
        \caption{LPIPS of SemSteDiff with DeepJSCC}
        \label{fig:deep+lpips}
    \end{subfigure}
    \caption{Performance of SemSteDiff scheme with DeepJSCC across four metrics. (a) shows the trend of PSNR under different trained SNRs and tested SNRs. (b), (c) and (d) show the trend of MSE, SSIM and LPIPS, respectively. }
    \label{figure:deepjscc}
    \vspace{1mm}
\end{figure*}

\begin{figure*}[ ]
    \centering
    \begin{subfigure}[b]{0.24\textwidth}
        \includegraphics[width=\linewidth]{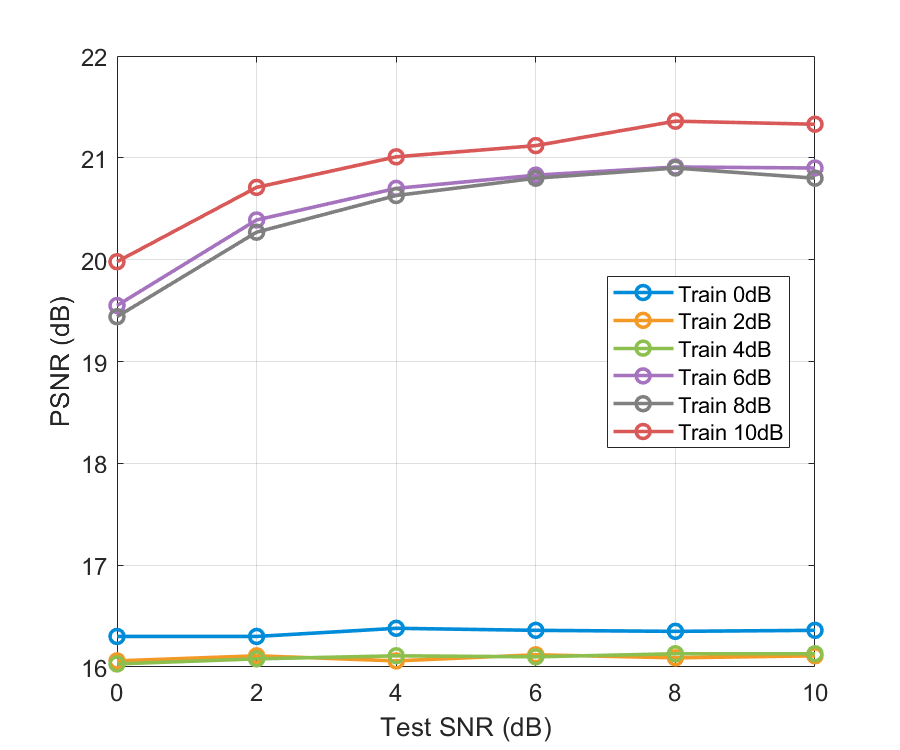}
        \caption{PSNR of SemSteDiff with NTSCC}
        \label{fig:ntscc+psnr}
    \end{subfigure}
    \hfill
    \begin{subfigure}[b]{0.24\textwidth}
        \includegraphics[width=\linewidth]{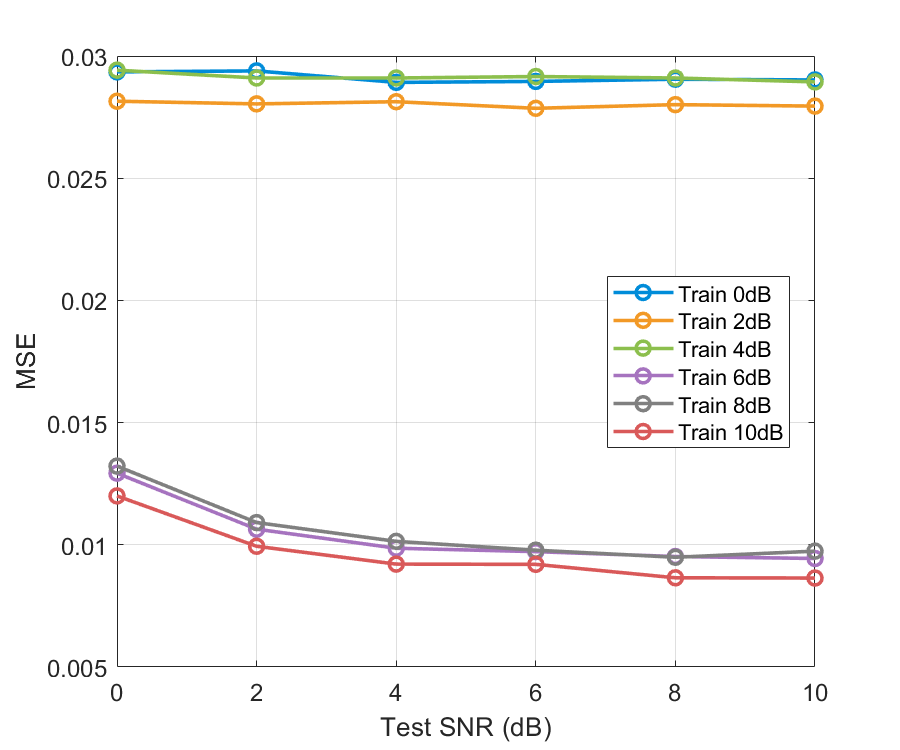}
        \caption{MSE of SemSteDiff with NTSCC}
        \label{fig:ntscc+mse}
    \end{subfigure}
    \hfill
    \begin{subfigure}[b]{0.24\textwidth}
        \includegraphics[width=\linewidth]{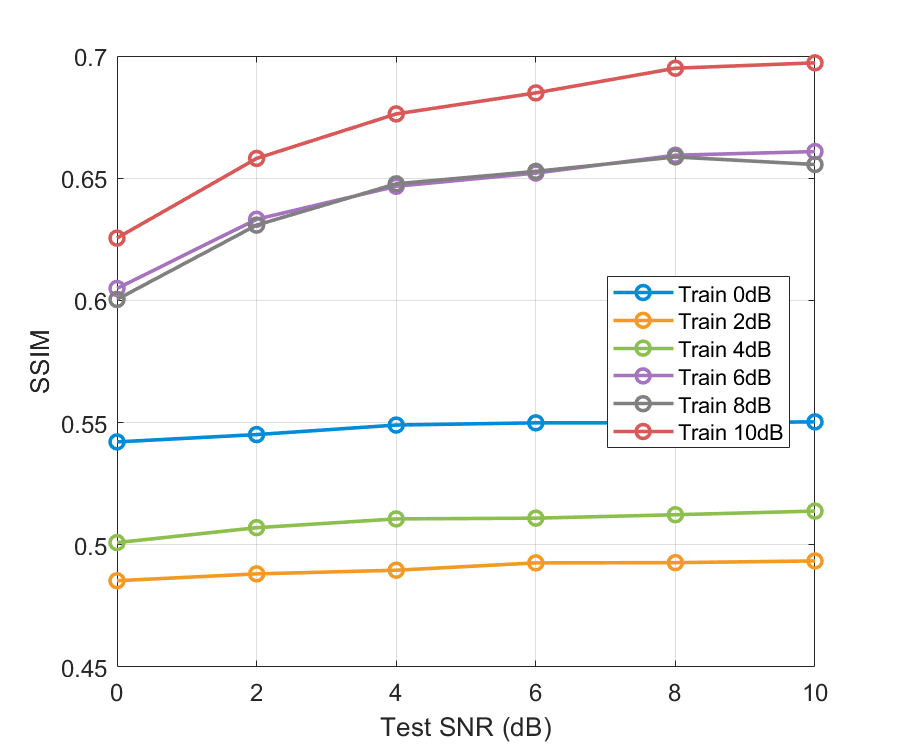}
        \caption{SSIM of SemSteDiff with NTSCC}
        \label{fig:ntscc+ssim}
    \end{subfigure}
    \hfill
    \begin{subfigure}[b]{0.24\textwidth}
        \includegraphics[width=\linewidth]{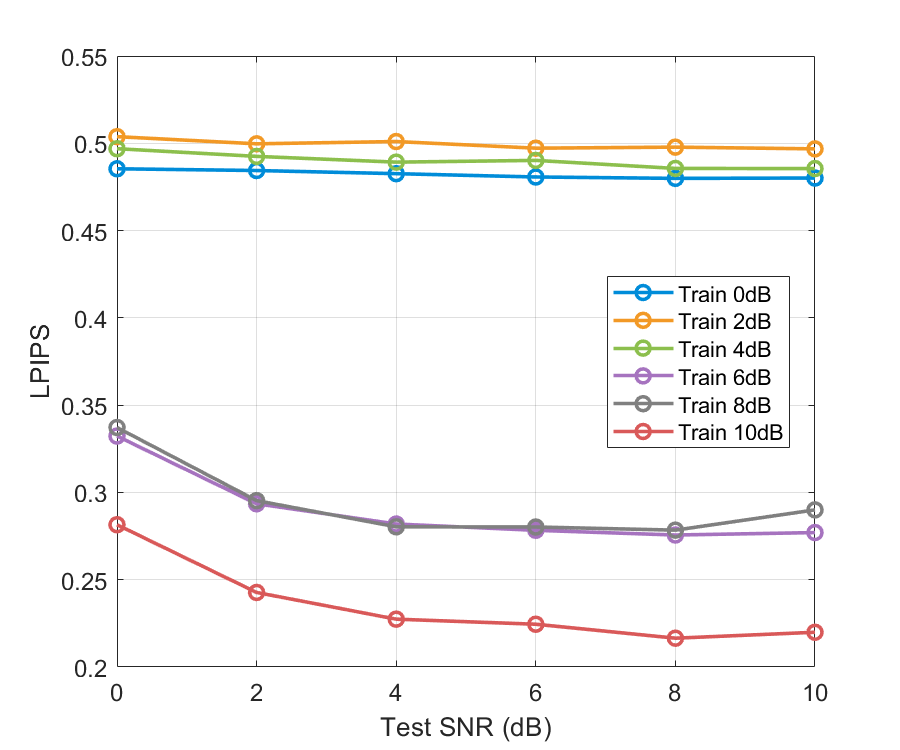}
        \caption{LPIPS of SemSteDiff with NTSCC}
        \label{fig:ntscc+lpips}
    \end{subfigure}
    \caption{Performance of SemSteDiff scheme with NTSCC across four metrics. (a) shows the trend of PSNR under different trained SNRs and tested SNRs. (b), (c) and (d) show the trend of MSE, SSIM and LPIPS, respectively. }
    \label{figure:ntscc}
    \vspace{1mm}
\end{figure*}

\begin{figure*}[]
    \centering
    \begin{subfigure}[b]{0.24\textwidth}
        \includegraphics[width=\linewidth]{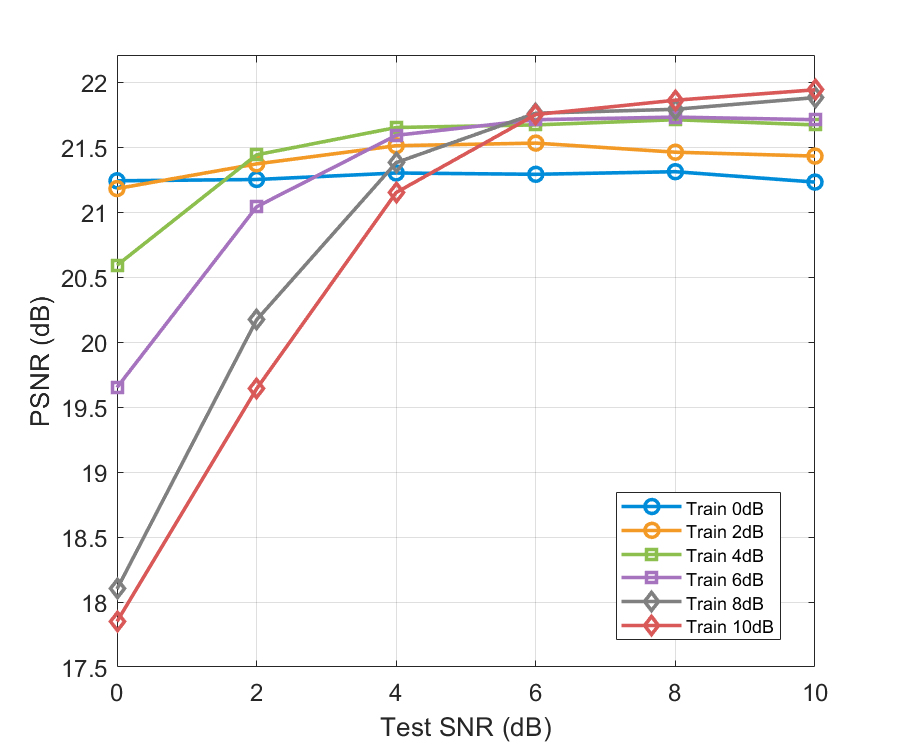}
        \caption{PSNR of SemSteDiff with SwinJSCC}
        \label{fig:swin+psnr}
    \end{subfigure}
    \hfill
    \begin{subfigure}[b]{0.24\textwidth}
        \includegraphics[width=\linewidth]{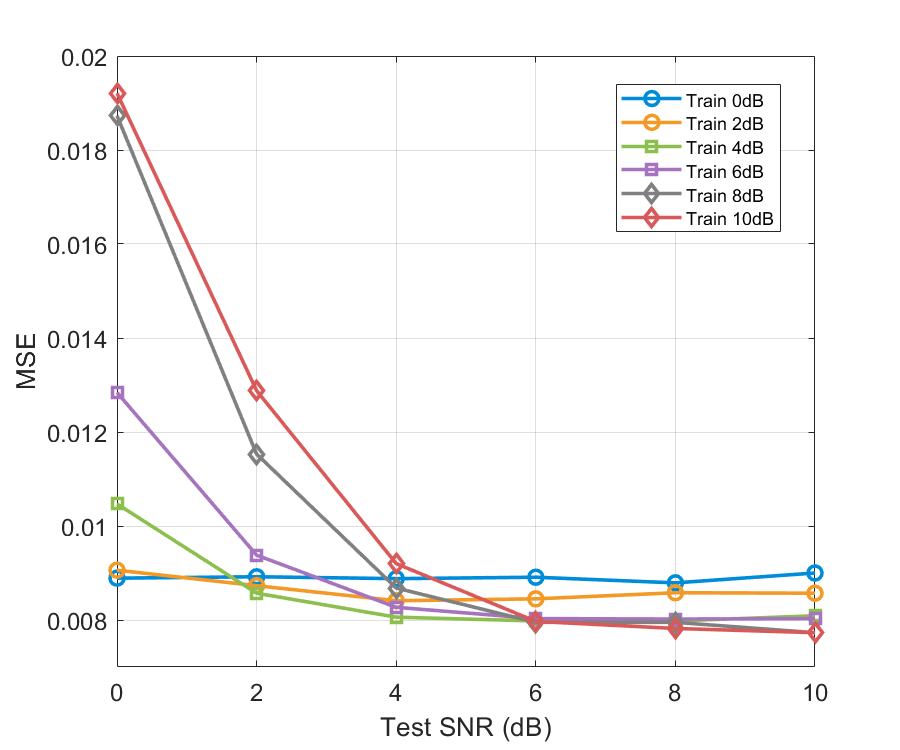}
        \caption{MSE of SemSteDiff with SwinJSCC}
        \label{fig:swin+mse}
    \end{subfigure}
    \hfill
    \begin{subfigure}[b]{0.24\textwidth}
        \includegraphics[width=\linewidth]{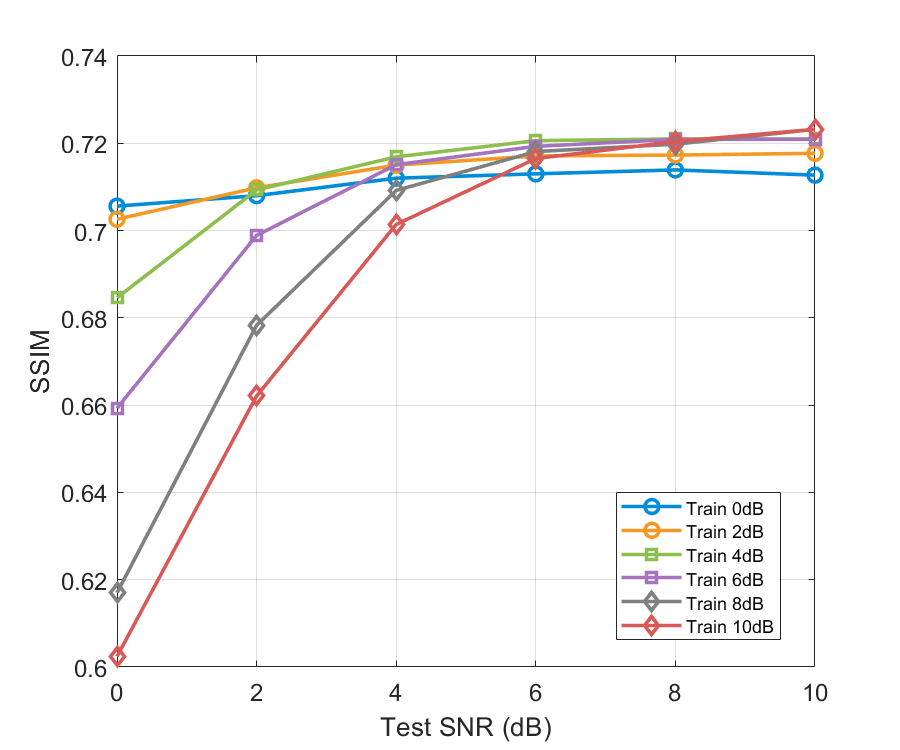}
        \caption{SSIM of SemSteDiff with SwinJSCC}
        \label{fig:swin+ssim}
    \end{subfigure}
    \hfill
    \begin{subfigure}[b]{0.24\textwidth}
        \includegraphics[width=\linewidth]{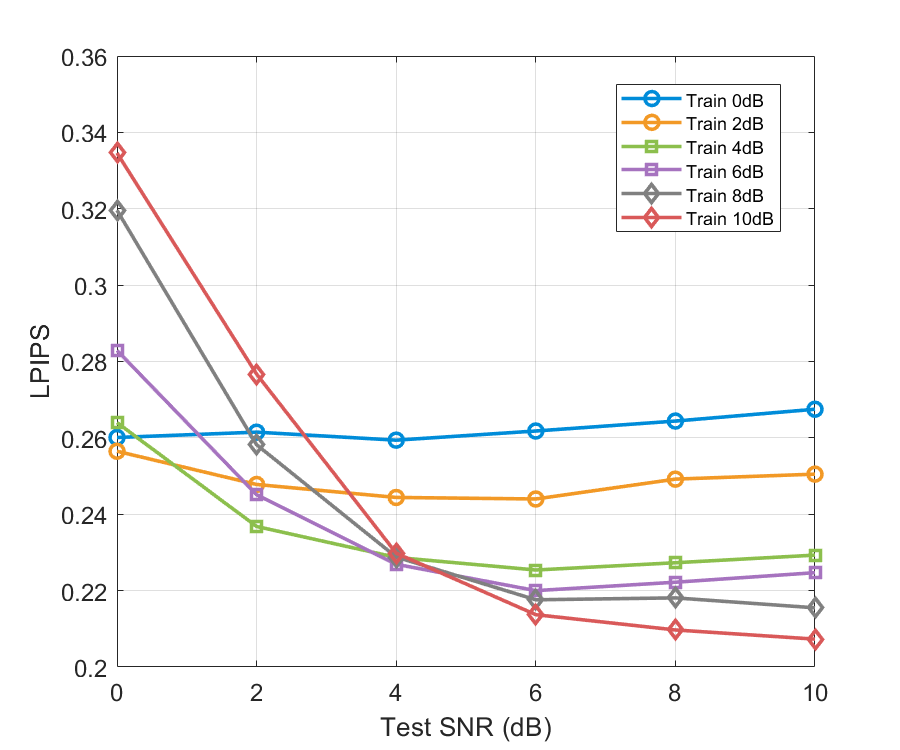}
        \caption{LPIPS of SemSteDiff with SwinJSCC}
        \label{fig:swin+lpips}
    \end{subfigure}
    \caption{Performance of SemSteDiff scheme with SwinJSCC across four metrics. (a) shows the trend of PSNR under different trained SNRs and tested SNRs. (b), (c) and (d) show the trend of MSE, SSIM and LPIPS, respectively. }
    \label{figure:swinjscc}
    \vspace{2mm}
\end{figure*}

\subsection{Simulation Results}
\subsubsection {Performance of the proposed scheme}
\begin{figure}[t]
\centering
\includegraphics[width=\linewidth]{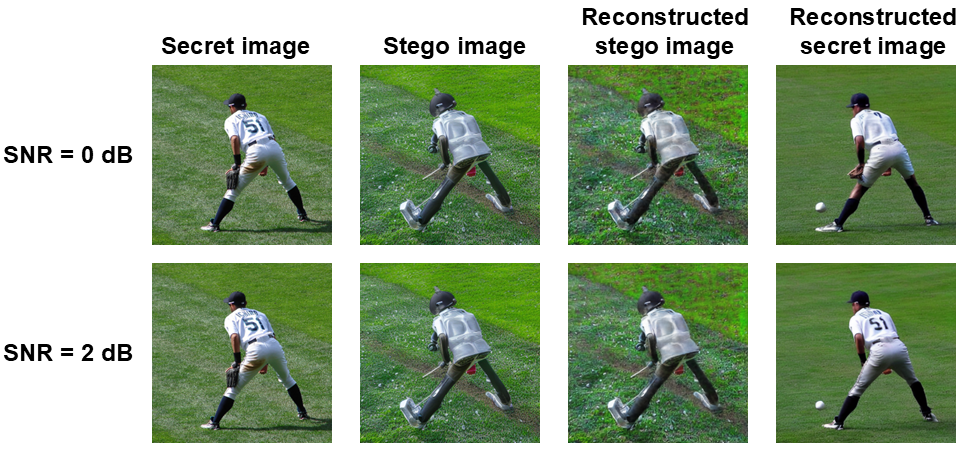}
\caption{Visualization of SemSteDiff on UniStega, where the private key is "a baseball player in a white uniform is bent over" and the public key is "a robot is bending over with his back facing the camera".}
\vspace{-5mm}
\label{unistega}
\end{figure}

As shown in Figure \ref{figure:deepjscc}, Figure \ref{figure:ntscc}, and Figure \ref{figure:swinjscc}, the proposed SemSteDiff framework consistently achieves robust performance in both distortion (PSNR and MSE) and perception (SSIM and LPIPS). In terms of distortion, PSNR exceeds 20 dB across most testing SNR values, with a peak of 21.94 dB with SwinJSCC-based SemSteDiff scheme under $\text{SNR}=10~\text{dB}$. Even when integrated with the lightweight DeepJSCC framework, SemSteDiff still achieves a PSNR of 21.20 dB. The trends in MSE also illustrate that SemSteDiff maintains a low reconstruction error, with the minimum MSE reaching 0.00773 for SwinJSCC and remaining below 0.0095 for DeepJSCC and NTSCC in most cases. These results demonstrate the effectiveness of the proposed scheme, which achieves low-distortion recovery even in the presence of poor channel environment. In terms of perception, SSIM scores consistently exceed 0.70, reaching up to 0.7232 in optimal configurations. Moreover, these perceptual similarities remain stable even under mismatching training and test SNRs. Also, LPIPS values remain uniformly low, with the best result of 0.2073 observed under the SwinJSCC framework. These perception results reflect SemSteDiff can preserve most semantic features, which is depended on the strong generation ability of diffusion model. Meanwhile, the analysis across different frameworks confirms that SemSteDiff is not depended on any specific JSCC architecture, but rather serves as a plug-and-play module. We further evaluate SemSteDiff in an additional dataset UniStega-Content \cite{yang2024diffstega}, and the experiment results are shown in Table \ref{tab:unistega} and Figure \ref{unistega}. According to Table \ref{tab:unistega}, SemSteDiff demonstrates stable performance under varying channel conditions. As SNR increases from 0 dB to 2 dB, the PSNR improves from 18.06 dB to 18.36 dB and the SSIM increases to 0.50, indicating consistent reconstruction reliability. The visual results in Figure \ref{unistega} further validate this robustness, showing that the ``baseball player" secret image is successfully recovered even at a low SNR of 0 dB, while the stego image effectively maintains the appearance of the "robot" public key despite severe channel interference. The above discussion reflects SemSteDiff's broad applicability and generalization for SemCom.

\begin{table}[t]
\centering
\caption{Performance of SemSteDiff on UniStega, where the semantic codec module is trained at 0 dB in DeepJSCC framework.}
\label{tab:unistega}
\small
\setlength{\tabcolsep}{6pt}
\renewcommand{\arraystretch}{1.1}
\begin{tabular}{c c c c c}
\toprule
\textbf{SNR (dB)} & \textbf{PSNR (dB)} & \textbf{SSIM} & \textbf{LPIPS} & \textbf{MSE} \\
\midrule
0  & 18.06 & 0.48 & 0.43 & 0.017 \\
2  & 18.36 & 0.50 & 0.42 & 0.016 \\
\bottomrule
\end{tabular}
\end{table}

To evaluate the importance of the private key, we simulate the scheme that the coverless SemSteDiff without private key as comparison. As shown in Table \ref{tab:ablation_private_key_psnr}, the PSNR of SemSteDiff is 19.78 dB but the comparison scheme only achieves 16.74 dB at $\text{SNR}=0~\text{dB}$, for private key preserves semantic information to reconstruct the secret image as an guarantee. The comparison scheme is theoretically capable of recovering the secret image based on the invertibility of DDIM, however, it fails to achieve robustness as SemSteDiff under the inference of channel noise and semantic compression. The design of dual semantic key ensures successful employment of coverless steganography in lossy transmission environment, which demonstrates the effective of SemSteDiff.

\begin{table}[t]
\centering
\caption{PSNR comparison of the proposed scheme with and without private key, where the DeepJSCC framework is trained at SNR = 10 dB.}
\label{tab:ablation_private_key_psnr}
\begin{tabular}{ccc}
\toprule
\textbf{SNR (dB)} 
& \textbf{Without Private Key} 
& \textbf{With Private Key} \\
\midrule
0  & 16.74 dB & 19.78 dB \\
2  & 17.29 dB & 20.42 dB \\
4  & 17.69 dB & 20.68 dB \\
6  & 18.00 dB & 20.98 dB \\
8  & 18.02 dB & 21.08 dB \\
10 & 18.10 dB & 21.20 dB \\
\bottomrule
\end{tabular}
\end{table}

\begin{figure*}[]
    \centering
    \begin{subfigure}[b]{0.24\textwidth}
        \includegraphics[width=\linewidth]{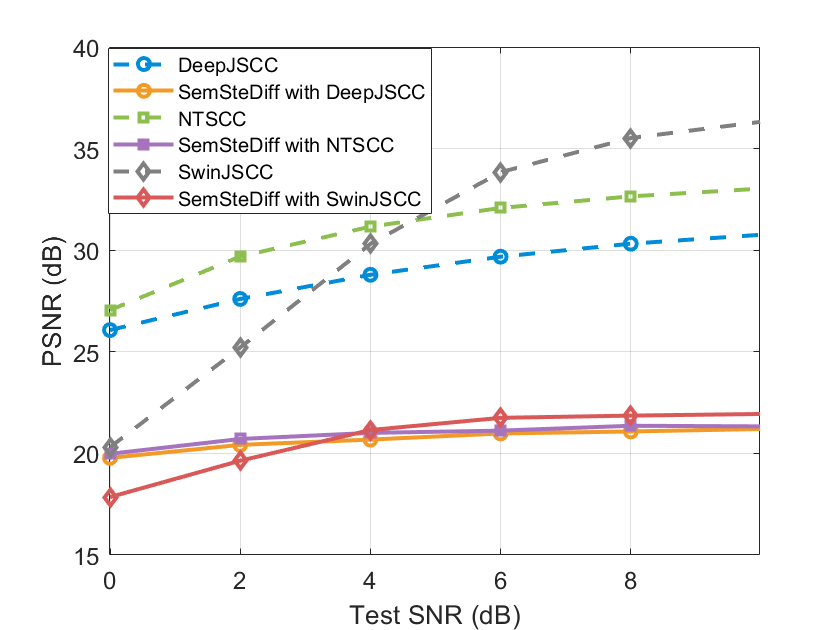}
        \caption{Comparison on PSNR}
        \label{comparePSNR}
    \end{subfigure}
    \hfill
    \begin{subfigure}[b]{0.24\textwidth}
        \includegraphics[width=\linewidth]{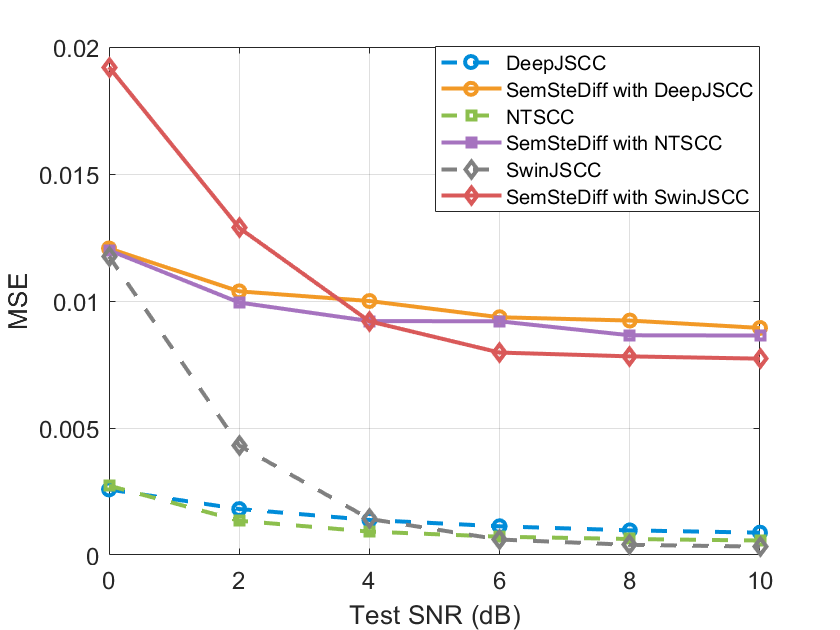}
        \caption{Comparison on MSE}
        \label{compareMSE}
    \end{subfigure}
    \hfill
    \begin{subfigure}[b]{0.24\textwidth}
        \includegraphics[width=\linewidth]{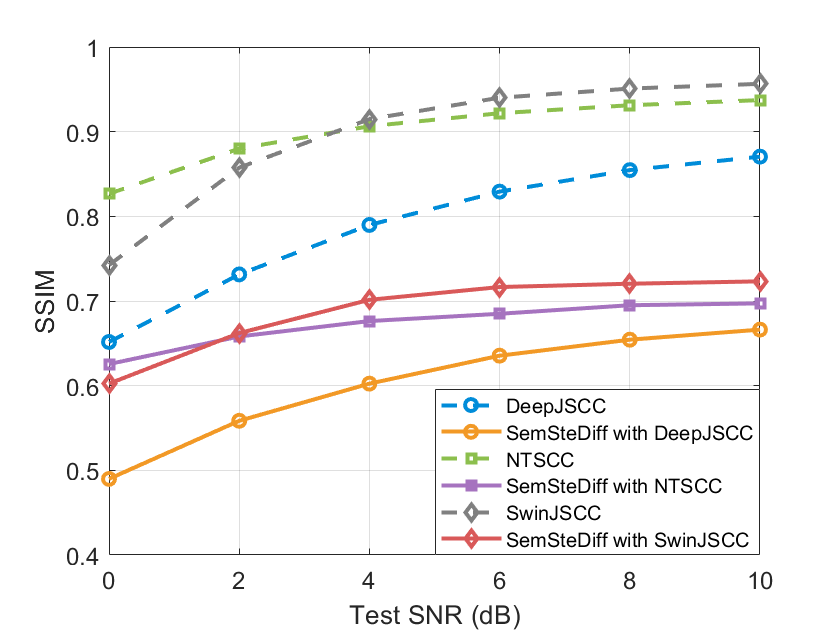}
        \caption{Comparison on SSIM}
        \label{compareSSIM}
    \end{subfigure}
    \hfill
    \begin{subfigure}[b]{0.24\textwidth}
        \includegraphics[width=\linewidth]{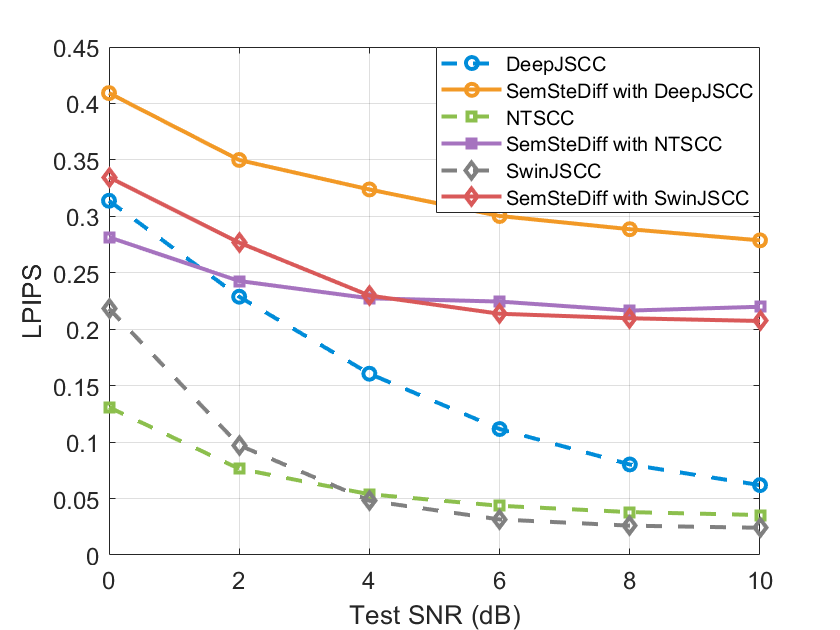}
        \caption{Comparison on LPIPS}
        \label{compareLPIPS}
    \end{subfigure}
    \caption{Comparison between SemSteDiff and baselines evaluated on four metrics. (a), (b), (c) and (d) show the trend of PSNR, MSE, SSIM and LPIPS under $\text{SNR}=10~\text{dB}$, respectively. The solid lines are the original JSCC simulation results, while the dash lines are the JSCC framework with SemSteDiff scheme. }
    \label{fig:compare}
\end{figure*}

Figure \ref{fig:compare} illustrates the comparison of whether employ our scheme to SemCom under $\text{SNR}=10~\text{dB}$. As expected, SemCom models without steganography present achieve notably higher PSNR and lower MSE. For instance, SwinJSCC without SemSteDiff attains peak PSNR of 36.32 dB and the MSE reaches 0.00033, while its counterpart with SemSteDiff achieves PSNR of 21.94 dB and MSE of 0.00773. Similar trends also appear in JSCC and NTSCC frameworks. This considerable gap in distortion metrics, however, does not imply a failure in reconstruction. It highlights a shift in the optimization target: SemSteDiff sacrifices pixel-level precision to achieve the target of security, enhancing the importance of semantic recovery and perceptual coherence. Under this paradigm, although JSCC coding parameters such as SNR primarily determine the distortion level of the transmitted stego image, the reconstruction quality of SemSteDiff is notably less sensitive to these variations compared to traditional SemCom models. As shown in Figure \ref{fig:compare}, higher SNR generally preserves more semantic features; however, the diffusion model provides a strong generative prior that partially compensates for channel-induced distortions. This results in a reduced performance gap across different SNRs, indicating the enhanced robustness of SemSteDiff against channel variations at the semantic level.
From this perspective, SSIM offers a more relevant measure of success. As we could find all SemSteDiff employments maintain SSIM values above 0.66 across the SNR range, indicating strong structural consistency. For instance, SwinJSCC-based SemSteDiff reaches 0.7232 at 10 dB, and even the DeepJSCC-based SemSteDiff achieves 0.6662, demonstrating the system's ability to retain layout and spatial relationships to human perceptions. LPIPS further confirms this trend for all simulations confirms LPIPS scores below 0.28. Moreover, SwinJSCC-based SemSteDiff achieves a minimum LPIPS of 0.2073, highlighting its alignment with human perception even under imperfect reconstruction conditions. 

\begin{figure}[]
\centering
\includegraphics[width=\linewidth]{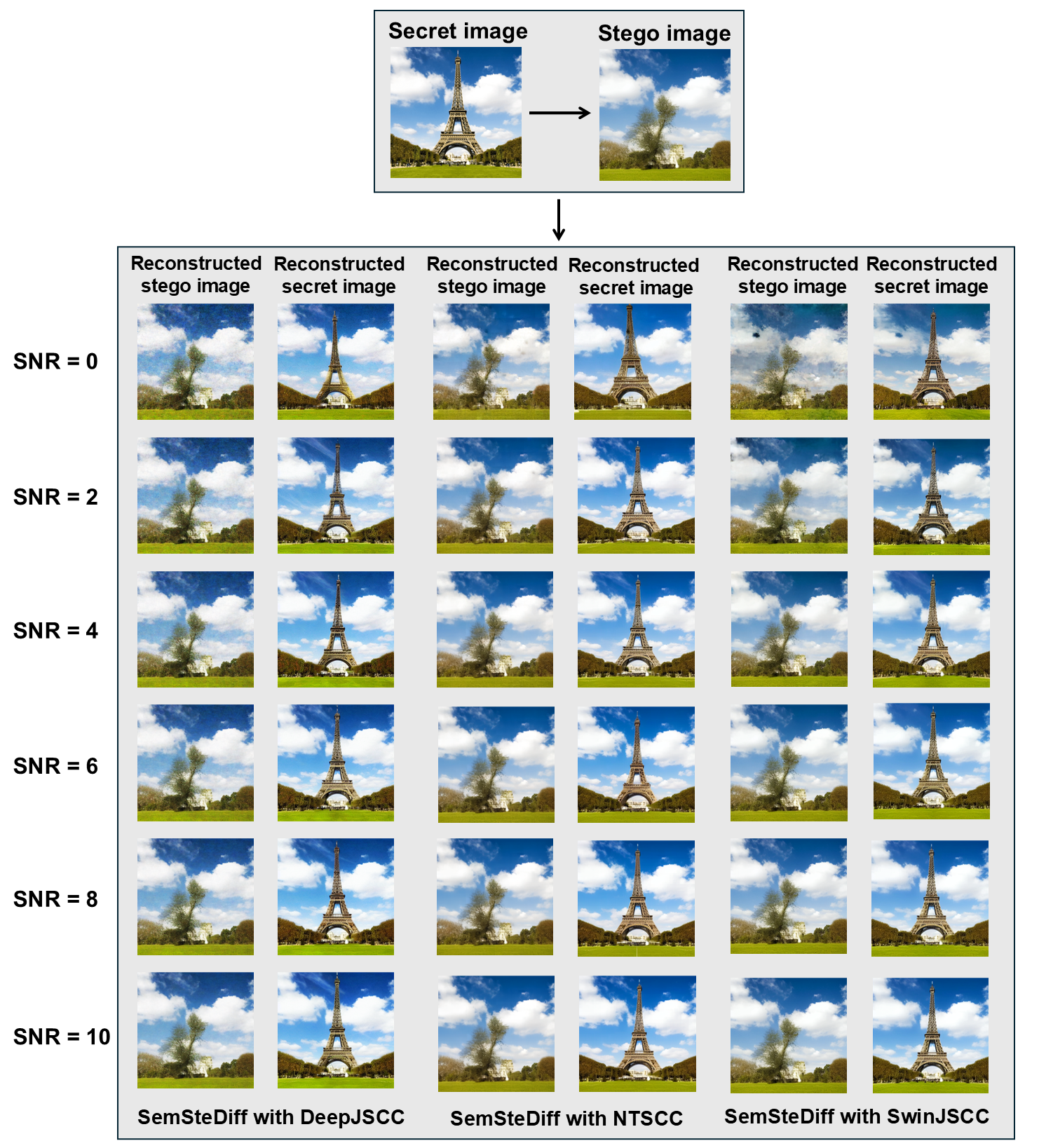}
\caption{Visualization of image steganography and reconstruction under different JSCC models, where the private key is ``an Eiffel Tower" and the public key is ``a tree". The top row shows the original secret image and its corresponding stego image, while the subsequent rows visualize the reconstructed stego and secret images under different SNRs in DeepJSCC, NTSCC, and SwinJSCC frameworks.}
\label{diffSNR}
\end{figure}

To intuitively validate the effect of the proposed SemSteDiff framework, Figure \ref{diffSNR} further presents the visualization of image steganography and reconstruction. The visualization results show that strong channel noise, especially under low SNR conditions (e.g., 0 dB), significantly distorts the reconstructed stego image, causing color drift, texture loss, and structural collapse. However, SemSteDiff demonstrates a remarkable ability to recover the semantic of the secret image, particularly for the main subject defined by the private key prompt. Even under poor conditions, critical visual features such as the outline, texture, and color of the target object (e.g. the Eiffel Tower) are preserved in the final reconstruction secret image. 
As a result, even when the reconstructed stego image fails to provide visually coherent cues, the diffusion model leverages the strong generation to reconstruct reasonable and meaningful visual representation of corresponding secret images. At higher SNR levels such as 10 dB, the overall reconstruction quality improves significantly across all JSCC models, with sharper edges, accurately restored colors, and better background consistency. 

\begin{figure*}[]
    \centering
    \begin{subfigure}[b]{0.24\textwidth}
        \centering
        \includegraphics[width=\linewidth]{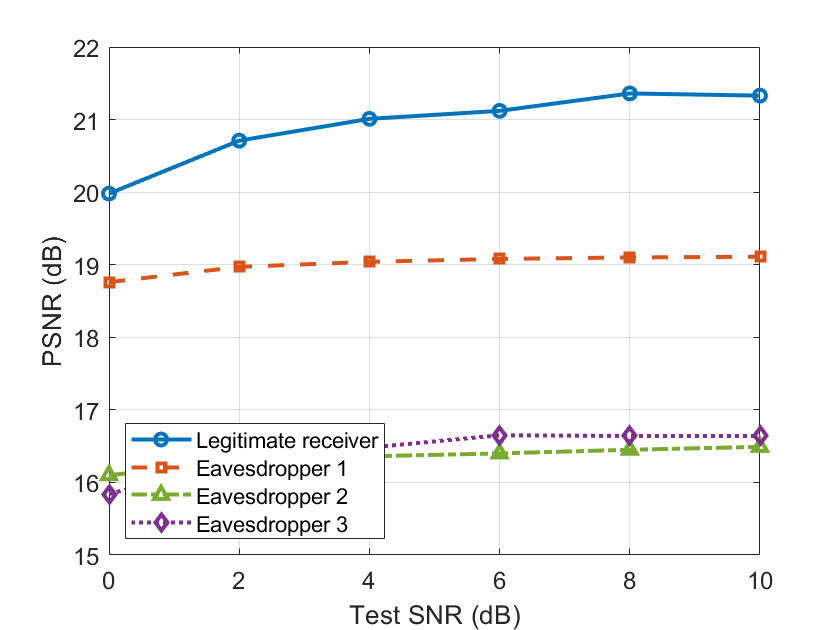}
        \caption{PSNR of eavesdroppers}
        \label{attackerPSNR}
    \end{subfigure}
    \hfill
    \begin{subfigure}[b]{0.24\textwidth}
        \centering
        \includegraphics[width=\linewidth]{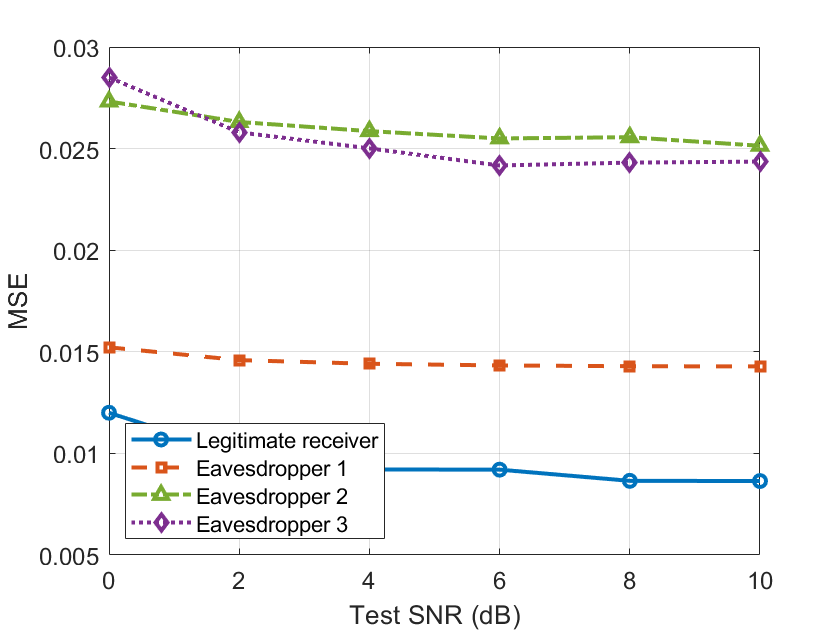}
        \caption{MSE of eavesdroppers}
        \label{attackerMSE}
    \end{subfigure}
    \hfill
    \begin{subfigure}[b]{0.24\textwidth}
        \centering
        \includegraphics[width=\linewidth]{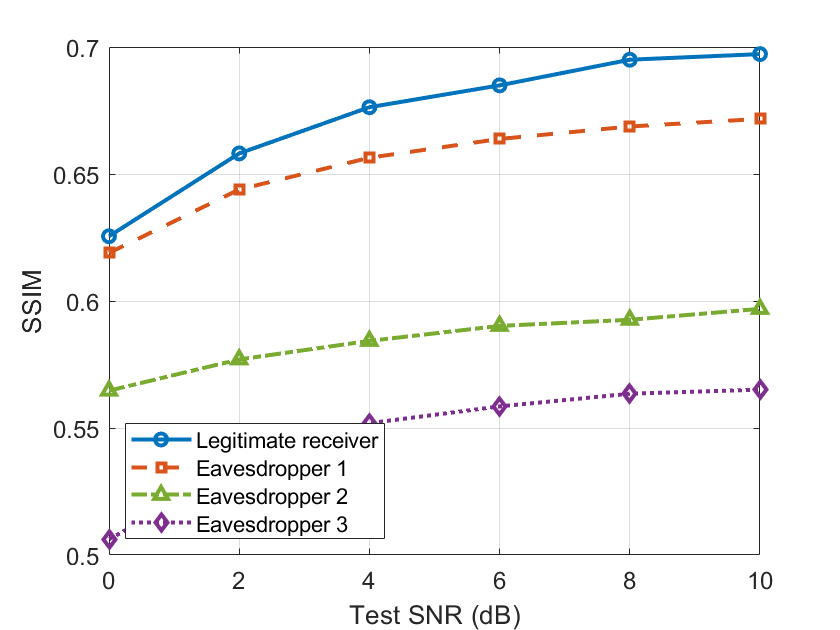}
        \caption{SSIM of eavesdroppers}
        \label{attackerSSIM}
    \end{subfigure}
    \hfill
    \begin{subfigure}[b]{0.24\textwidth}
        \centering
        \includegraphics[width=\linewidth]{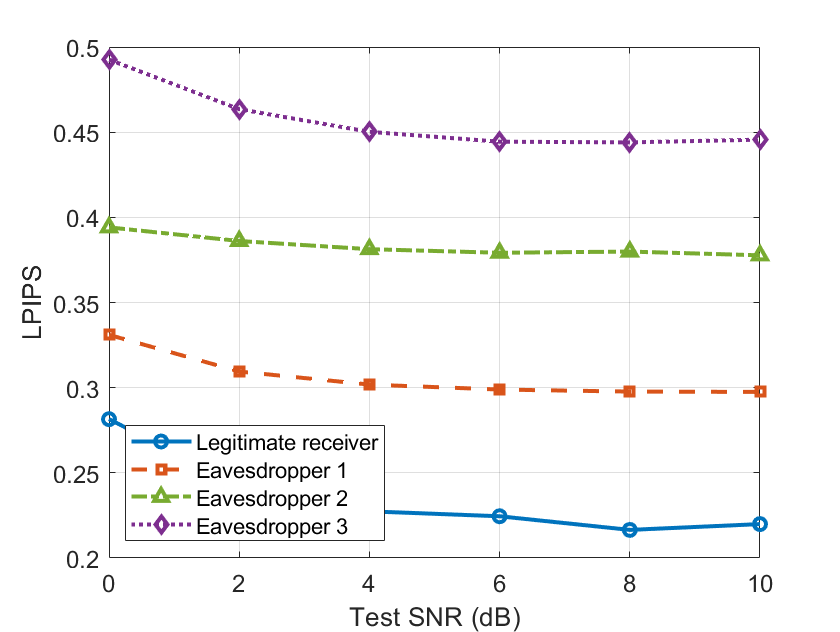}
        \caption{LPIPS of eavesdroppers}
        \label{attackerLPIPS}
    \end{subfigure}

    \caption{Comparison between legitimate receiver and eavesdroppers on four metrics. (a), (b), (c) and (d) show the comparison of PSNR, MSE, SSIM and LPIPS under $\text{SNR}=8~\text{dB}$, respectively. The solid line shows the performance of legitimate receiver, while the dash lines show the performance of three eavesdroppers. }
    \label{fig:attacker}
\end{figure*}

\begin{figure}[]
\centering
\includegraphics[width=\linewidth]{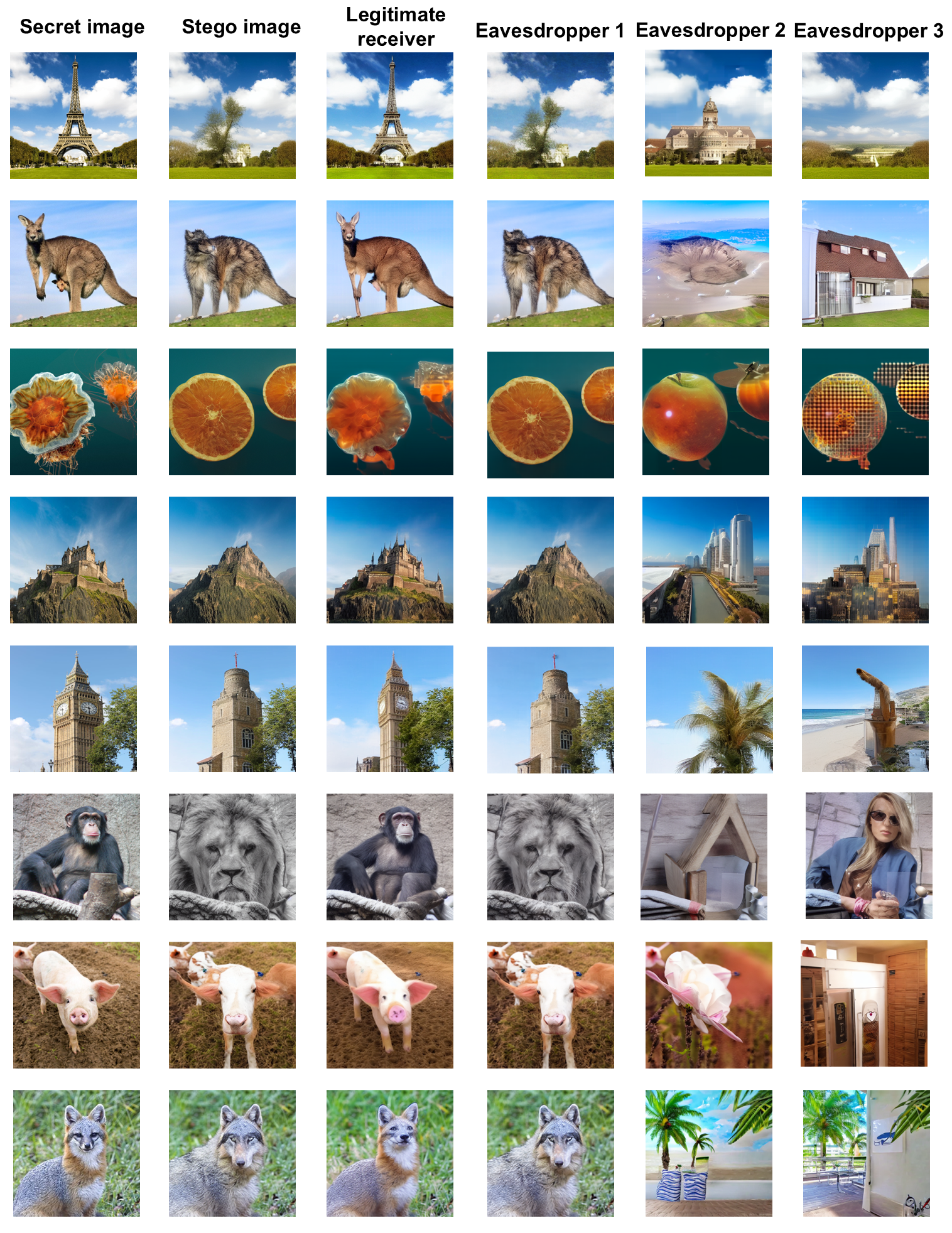}
\caption{Visualization of image steganography and reconstruction of legitimate receiver and eavesdroppers. The first column shows the secret image. The second column shows the stego image. The following four columns show the reconstructed images of legitimate receiver and eavesdroppers, respectively.}
\label{fig:mulImage}
\end{figure}

\subsubsection {Security evaluation under eavesdropping scenarios}
To further assess the security performance, we consider three eavesdroppers with different eavesdropping abilities are considered.

\begin{itemize}
\item \textbf{Eavesdropper 1:} Eavesdropper only possesses the semantic decoder without any key information. 
\item \textbf{Eavesdropper 2:} Eavesdropper possesses the semantic decoder and the correct public key, but the private key is wrong.
\item \textbf{Eavesdropper 3:} Eavesdropper possesses the semantic decoder and the correct public key, but without the private key.
\end{itemize}

Figure \ref{fig:attacker} shows the quantified performance of SemSteDiff with legitimate receiver and eavesdroppers n SNR = 8 dB of NTSCC framework. The legitimate receiver performs much better than eavesdroppers through all metrics, demonstrating strong semantic protection. Eavesdropper 1, which has no access to either key, simply intercepts the reconstructed stego image. It leads to the SSIM around 0.6717 and PSNR around 19.11 dB, but LPIPS remains high at 0.3, which shows similar structural of secret image but different semantic meaning. This phenomenon is caused by our key design strategy: the public key and private key are paraphrased to be semantically distinct but visually similar, enabling the diffusion model to generate natural-looking images to confuse the eavesdroppers. Eavesdropper 2 performs significantly worse with PSNR drops to 16.49 dB, SSIM to 0.5969, and LPIPS increases to 0.3777. It only has a correct public key but an incorrect private key, causing the output shifts entirely away from the secret image. Eavesdropper 3 with no private key performance similar in metric as eavesdropper 2. Although the public key provides some semantic information, the absence of private key guidance results in the generated images retaining only the general contours and posture similar to the secret images, the generated content is completely unrelated.
Moreover, the secure performance is also influenced by SNR, which determines the amount of reliable semantic information available in the transmitted stego image. As the test SNR increases from 0 dB to 10 dB, the legitimate receiver's PSNR improves significantly from 19.98 dB to 21.33 dB, whereas Eavesdropper 1 only sees a marginal rise from 18.76 dB to 19.11 dB. This widening gap reveals that legitimate receivers leverage semantic keys to guide the reverse diffusion process could defend the channel noise, while eavesdroppers lacking such guidance or wrong guidance cannot compensate for this degradation. The influence in JSCC parameters tend to enlarge the semantic gap between legitimate receivers and eavesdroppers, reinforcing the overall security of SemSteDiff.


Figure \ref{fig:mulImage} illustrates the visualization of a subset of images reconstruction with legitimate receiver and eavesdroppers in SNR = 8 dB of NTSCC framework. The legitimate receiver is able to reconstruct images that preserve both structure and semantics across all categories, including landmarks, animals, natural scenes, and etc. However, the eavesdroppers consistently fail to infer the correct content. Notably, even when visual textures and contours appear similar, the semantic essence is either distorted or entirely incorrect in eavesdropper outputs, demonstrating that the generative process cannot be easily manipulated without access to both semantic keys. Consider the sixth row of Figure \ref{fig:mulImage}, the private key is “chimpanzee”; the corresponding public key is "lion". The legitimate receiver reconstructs the secret image correctly. In contrast, Eavesdropper 1 just recovers the stego image of a lion. Eavesdropper 2 only has correct public key but an incorrect private key “cabin”. In this case, the generated image shows a content of architecture, completely misaligned with the chimpanzee. Eavesdropper 3 without private key only generates an unrelated image of a person with similar posture as chimpanzee. These results highlight the robustness and confidentiality of our SemSteDiff framework. By ensuring that both public and private keys are indispensable for reconstructing secret images, the system effectively prevents unauthorized access, even in scenarios where partial key information is leaked. 
Moreover, the visual diversity of the selected examples emphasizes that this security property is not limited to specific visual patterns, but persists across various inputs, further validating the reliability of our approach under diverse threat conditions.
\begin{table}[]
\centering
\caption{Comparison between legitimate receiver and eavesdroppers on CLIP score, where the semantic codec module is trained at 8 dB in NTSCC framework}
\label{tab:clip_score_snr}
\setlength{\tabcolsep}{7pt}
\renewcommand{\arraystretch}{1.1}
\begin{tabular}{c c c c c}
\toprule
\textbf{SNR (dB)} & \textbf{Legit} & \textbf{Eaves1} & \textbf{Eaves2} & \textbf{Eaves3} \\
\midrule
0  & 28.36 & 24.83 & 17.87 & 18.07 \\
2  & 28.57 & 24.87 & 18.41 & 18.39 \\
4  & 28.58 & 24.82 & 18.71 & 19.26 \\
6  & 28.27 & 24.75 & 18.91 & 19.56 \\
8  & 28.34 & 24.63 & 19.07 & 19.67 \\
10 & 28.16 & 24.68 & 19.09 & 19.80 \\
\bottomrule
\end{tabular}
\end{table}

\begin{table}[]
\centering
\caption{Steganalysis detection accuracy of XuNet \cite{xu2016structural}.}
\label{tab:xunet}
\begin{tabular}{l c}
\toprule
\textbf{Methods} & \textbf{Detection Accuracy (\%)} \\
\midrule
Baluja\cite{baluja2019hiding} & 95.12\\
ISN\cite{lu2021large}& 56.23 \\
HiNet\cite{jing2021hinet} & 55.71 \\
SemSteDiff & 52.18 \\
\bottomrule
\end{tabular}
\end{table}

Table~\ref{tab:clip_score_snr} presents the CLIP-based semantic consistency between the reconstructed images and the secret keys under different SNRs. It can be observed that the legitimate receiver consistently achieves the highest CLIP scores across all SNR values, with scores around 28.2 to 28.6, while all eavesdroppers exhibit significantly lower semantic similarity. These results confirm that SemSteDiff achieves semantic-level security, where meaningful semantic information can be reliably recovered by legitimate users while being effectively concealed from unauthorized eavesdroppers.
Furthermore, Table \ref{tab:xunet} shows the steganalysis detection accuracy of representative traditional image steganography methods integrated with the same NTSCC transmission framework. It can be observed that conventional cover-modification-based methods, including Baluja \cite{baluja2019hiding}, ISN\cite{lu2021large}, and HiNet\cite{jing2021hinet}, exhibit higher detection accuracy under XuNet\cite{xu2016structural}. In contrast, the proposed SemSteDiff achieves detection accuracy close to random guessing, demonstrating strong resistance against DL-based steganalysis. This comparison verifies that avoiding explicit cover modification could improve covertness of steganography, thereby enhancing the security of semantic communication under eavesdropping threats.

\begin{figure}[]
\centering
\includegraphics[width=\linewidth]{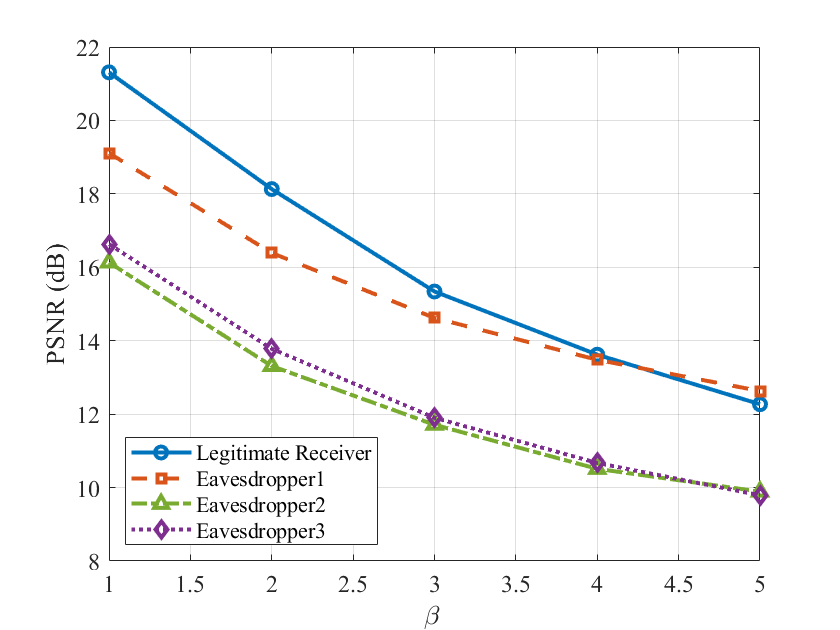}
\caption{Trade-off between guidance scale $\beta$ and PSNR, where the semantic codec module is trained at 10 dB in NTSCC framework}
\label{guidance}
\vspace{-5mm}
\end{figure}

Figure \ref{guidance} further investigates the impact of guidance scale $\beta$ in (\ref{30}) on the security–performance trade-off of SemSteDiff. As $\beta$ increases, the diffusion process is more strongly constrained by the conditioning semantic key, leading to improved resistance against eavesdroppers while maintaining acceptable reconstruction fidelity at the legitimate receiver. However, when $\beta$ exceeds 2, the reconstruction PSNR of the legitimate receiver drops rapidly from 18.13 dB at $\beta=2$ to 12.27 dB at $\beta=5$, whereas the security gain becomes marginal, indicating a diminishing return region where stronger guidance sacrifices communication performance without proportional security improvement. Also, a larger $\beta$ enforces a tighter alignment to semantic keys. The keys only contains general description while limiting semantic detail in diffusion process. The tighter alignment could degrades the availability for preserving specific private details by reducing the entropy of the sampling distribution, which focus the the generative trajectory on a narrow semantic key.

\section{Conclusions}
In this paper, we have introduced SemSteDiff to achieve coverless SemSteCom. Based on diffusion model, SemSteCom employs a semantic-key controlled conditional latent diffusion process to generate coverless stego image. Only legitimate receivers with both private and public keys enable to decode the correct secret image. Experimental results have demonstrated its effectiveness in guaranteeing semantic recovery of secret image for legitimate receivers and defending against intelligent eavesdroppers.  In the future, we will devote to mathematical derivation of the semantic steganographic capacity, providing systematically definition and capacity boundaries.
Meanwhile, we will further consider using lightweight LLMs fine-tuned specifically for key generation, or applying knowledge distillation to extract a task-specific lightweight model from a large-scale LLM, to improve the deployment consumption. Adopting advanced sampling method such as one-step-sampling is also put into consideration to reduce the generation latency. 
Furthermore, by jointly optimizing both stego image generation and semantic transmission, we will focus on developing an end-to-end scheme to achieve lightweight and effective coverless semantic steganography communication.

\bibliographystyle{IEEEtran}
\bibliography{ref}

\end{document}